\documentclass[a4paper,11pt]{article}

\usepackage[labelfont=bf,labelsep=period]{caption}
\usepackage[nodayofweek]{datetime}
\usepackage{enumitem}
\usepackage{hyperref}
\usepackage{float}
\usepackage[margin=2.5cm]{geometry}
\usepackage{graphicx}
\usepackage{numbertabbing}
\usepackage{times}
\usepackage{url}
\usepackage{xspace}
\usepackage{algorithm}
\usepackage{mathtools}
\usepackage{xcolor}
\usepackage{dirtytalk}
\usepackage{booktabs}
\usepackage{siunitx}
\usepackage{tabularx}
\usepackage{stackrel}
\usepackage{amsmath}

\setlist[]{topsep=2pt,partopsep=2pt,parsep=2pt,itemsep=2pt}

\floatstyle{ruled}
\newfloat{algo}{htbp}{algo}
\floatname{algo}{Algorithm} 

\usepackage{amsmath} 
\allowdisplaybreaks[2]        
 
\usepackage{amssymb} 

\usepackage{amsthm} 
\newtheoremstyle{plain-boldhead}
  {\topsep}
  {\topsep}
  {\itshape}
  {}
  {\bfseries}
  {.}
  { }
  {\thmname{#1}\thmnumber{ #2}\thmnote{ (\bfseries #3)}}
\newtheoremstyle{definition-boldhead}
  {\topsep}
  {\topsep}
  {\normalfont}
  {}
  {\bfseries}
  {.}
  { }
  {\thmname{#1}\thmnumber{ #2}\thmnote{ (\bfseries #3)}}
\theoremstyle{plain-boldhead}
\newtheorem{theorem}{Theorem}

\newtheorem{lemma}[theorem]{Lemma}

\theoremstyle{definition-boldhead}
\newtheorem{definition}{Definition}

\floatstyle{ruled}
\newfloat{algo}{htbp}{algo}
\floatname{algo}{Algorithm}

\renewcommand{\P}{\mathrm{P}}
\newcommand{\E}{\mathrm{E}}
\newcommand{\str}[1]{\textsc{#1}}
\newcommand{\var}[1]{\textit{#1}}
\newcommand{\op}[1]{\textsl{#1}}
\newcommand{\msg}[2]{\ensuremath{\ifempty{#2} [\str{#1}] \else [\str{#1}, {#2}] \fi}}

\newcommand{\false}{\textsc{false}\xspace}
\newcommand{\true}{\textsc{true}\xspace}

\newcommand{\CA}{\ensuremath{\mathcal{A}}\xspace}

\newcommand{\CD}{\ensuremath{\mathcal{D}}\xspace}

\newcommand{\CF}{\ensuremath{\mathcal{F}}\xspace}
\newcommand{\CG}{\ensuremath{\mathcal{G}}\xspace}

\newcommand{\CK}{\ensuremath{\mathcal{K}}\xspace}

\newcommand{\CN}{\ensuremath{\mathcal{N}}\xspace}

\newcommand{\CP}{\ensuremath{\mathcal{P}}\xspace}
\newcommand{\CQ}{\ensuremath{\mathcal{Q}}\xspace}
\newcommand{\CR}{\ensuremath{\mathcal{R}}\xspace}
\newcommand{\CS}{\ensuremath{\mathcal{S}}\xspace}
\newcommand{\CT}{\ensuremath{\mathcal{T}}\xspace}

\newcommand{\CV}{\ensuremath{\mathcal{V}}\xspace}
\newcommand{\CVF}{\ensuremath{\mathcal{VF}}\xspace}

\def \ifempty#1{\def\temp{#1} \ifx\temp\empty }

\newcommand\para[1]{\paragraph{#1.}}

\begin{document}

\title{\bf When is Spring coming? A Security Analysis of Avalanche Consensus}

\author{Ignacio Amores-Sesar\\
  University of Bern\footnote{Institute of Computer Science, University of Bern,
  Neubr\"{u}ckstrasse 10, 3012 CH-Bern, Switzerland.}\\
  \url{ignacio.amores@unibe.ch} 
  \and Christian Cachin\\
  University of Bern\footnotemark[1]\\
  \url{christian.cachin@unibe.ch}
  \and Enrico Tedeschi\\
  The Artic University of Norway\footnote{Institute of Computer Science, The Artic University of Norway,
  Hansine Hansens vei 54, 6050 NO-Langnes, Norway.}\\
  \url{enrico.tedeschi@uit.no}
}

\date{\today}

\maketitle

\begin{abstract}\noindent
  Avalanche is a blockchain consensus protocol with exceptionally low
  latency and high throughput.  This has swiftly established the
  corresponding token as a top-tier cryptocurrency.  Avalanche achieves
  such remarkable metrics by substituting proof of work with a random
  sampling mechanism.  The protocol also differs from Bitcoin, Ethereum,
  and many others by forming a directed acyclic graph (DAG) instead of a
  chain.  It does not totally order all transactions, establishes a partial
  order among them, and accepts transactions in the DAG that satisfy
  specific properties.  Such parallelism is widely regarded as a technique
  that increases the efficiency of consensus.

  Despite its success, Avalanche consensus lacks a complete abstract
  specification and a matching formal analysis.  To address this drawback,
  this work provides first a detailed formulation of Avalanche through
  pseudocode.  This includes features that are omitted from the original
  whitepaper or are only vaguely explained in the documentation.  Second,
  the paper gives an analysis of the formal properties fulfilled by
  Avalanche in the sense of a generic broadcast protocol that only orders
  related transactions.  Last but not least, the analysis reveals a
  vulnerability that affects the liveness of the protocol.  A possible
  solution that addresses the problem is also proposed.
\end{abstract}

\section{Introduction}
\label{sec:intro}

The Avalanche blockchain with its fast and scalable consensus protocol is
one of the most prominent alternatives to first-generation networks like
Bitcoin and Ethereum that consume huge amounts of energy.  Its AVAX token
is ranked 14th according to market capitalization in August
2022~\cite{coinmarketcap}. Avalanche offers a protocol with high
throughput, low latency, excellent scalability, and a lightweight
client. In contrast to many well-established distributed ledgers, Avalanche
is not backed by proof of work. Instead, Avalanche bases its security on a
deliberately metastable mechanism that operates by repeatedly sampling the
network, guiding the honest parties to a common output. This allows
Avalanche to reach a peak throughput of up to 20'000 transactions per
second with a latency of less than half a second~\cite{avalanche}.

This novel mechanism imposes stricter security constraints on Avalanche
compared to other networks.  Traditional Byzantine fault-tolerant consensus
tolerates up to a third of the parties to be
corrupted~\cite{DBLP:journals/jacm/PeaseSL80} and proof-of-work protocols
make similar assumptions in terms of mining
power~\cite{DBLP:conf/eurocrypt/GarayKL15,DBLP:conf/fc/EyalS14}.
Avalanche, however, can tolerate only up to $O(\sqrt{n})$ malicious
parties.  Furthermore, the transactions in the ``exchange chain'' of
Avalanche (see below) are not totally ordered, in contrast to most other
cryptocurrencies, which implement a form of atomic
broadcast~\cite{DBLP:books/daglib/0025983}.  As the protocol is structured
around a directed acyclic graph (DAG) instead of a chain, it permits some
parallelism.  Thus, the parties may output the same transactions in a
different order, unless these transactions causally depend on each other.
Only the latter must be ordered in the same way.

The consensus protocol of a blockchain is of crucial importance for its
security and for the stability of the corresponding digital assets.
Analyzing such protocols has become an important topic in current research.
Although Bitcoin appeared first without formal arguments, its security has
been widely understood and analyzed meanwhile.  The importance of proving
the properties of blockchain protocols has been recognized for a long
time~\cite{DBLP:conf/wdag/CachinV17}.

However, there are still protocols released today without the backing of
formal security arguments. The Avalanche whitepaper~\cite{avalanche}
introduces a family of consensus protocols and offers rigorous security
proofs for some of them. Yet the Avalanche protocol itself and the related
Snowman protocol, which power the platform, are not analyzed.  Besides,
several key features of this protocol are either omitted or described only
vaguely.

In this paper, we explain the Avalanche consensus protocol in detail.  We
describe it abstractly through pseudocode and highlight features that may
be overlooked in the whitepaper
(Sections~\ref{sec:model}--\ref{sec:description}).  Furthermore, we use our
insights to formally establish safety properties of Avalanche.  Per
contra, we also identify a weakness that affects its liveness.  In
particular, Avalanche suffers from a vulnerability in how it accepts
transactions that allows an adversary to delay targeted transactions by
several orders of magnitude (Section~\ref{sec:analysis}), which may render
the protocol useless in practice.
The problem results from dependencies that exist
among the votes on different transactions issued by honest parties; the
whitepaper does not address them.  The attack may be mounted by a single
malicious party with some insight into the network topology.
Finally, we suggest a modification to the Avalanche protocol that would
prevent our attacks from succeeding and reinstantiate liveness of the
protocol (Section~\ref{app:glacier}).  This version, which we call
\emph{Glacier}, restricts the sampling choices in order to break the
dependencies, but also eliminates the parallelism featured by Avalanche.

The vulnerability has been acknowledged by the Avalanche developers.  The
deployed version of the protocol differes however from the protocol in the
whitepaper in a crucial way.  It implements another measure that prevents
the problem, as we explain as well (in Appendix~\ref{appendix:fix}).

\section{Related work}

Despite Avalanche's tremendous success, there is no independent research on
its security.  Recall that Avalanche introduces the ``snow family'' of
consensus protocols based on sampling~\cite{avalanche,documentation}:
Slush, Snowflake, and Snowball.  Detailed proofs about liveness and safety
for the snow-family of algorithms are given.  The Avalanche protocol for
asset exchange, however, lacks such a meticulous analysis.  The
dissertation of Yin~\cite{DBLP:phd/us/Yin21} describes Avalanche as well,
but does not analyze its security in more detail either.

Recall that Nakamoto introduced Bitcoin~\cite{bitcoin} without any formal
analysis.  This has been corrected by a long line of research, which
established the conditions under which it is secure (e.g., by Garay,
Kiayias, and
Leonardos~\cite{DBLP:conf/eurocrypt/GarayKL15,DBLP:conf/crypto/GarayKL17}
and by Eyal and Sirer~\cite{DBLP:conf/fc/EyalS14}).

The consensus mechanisms that stand behind the best-known cryptocurrencies
are meanwhile properly understood.  Some of them, like the proof-of-stake
protocols of Algorand~\cite{DBLP:conf/sosp/GiladHMVZ17} and the Ouroboros
family that powers the Cardano
blockchain~\cite{DBLP:conf/crypto/KiayiasRDO17,DBLP:conf/eurocrypt/DavidGKR18},
did apply sound design principles by first introducing and analyzing the
protocols and only later implementing them.

Many others, however, have still followed the heuristic approach: they
released code first and were confronted with concerns about their security
later.  This includes
Ripple~\cite{DBLP:conf/trust/ArmknechtKMYZ15,DBLP:conf/opodis/Amores-SesarCM20}
and NEO~\cite{DBLP:conf/fc/WangYPB0D020}, in which several vulnerabilities
have been found, or Solana, which halted multiple times in 2021--2022.
Stellar comes with a formal model~\cite{DBLP:conf/sosp/LokhavaLMHBGJMM19},
but it has also been criticized~\cite{DBLP:conf/eurosp/KimKK19}.

Protocols based on DAGs have potentially higher throughput than those based
on chains.  Notable examples include PHANTOM and
GHOSTDAG~\cite{DBLP:conf/aft/SompolinskyWZ21}, the Tangle of
IOTA~(\url{www.iota.org}),
Conflux~\cite{DBLP:conf/usenix/LiLZYWYXLY20},
and others~\cite{DBLP:conf/podc/KeidarKNS21}.
However, they are also more complex to understand and 
susceptible to a wider range of attacks than those that use
a chain.  Relevant examples of
this kind are the IOTA protocol~\cite{DBLP:journals/corr/abs-2111-07805},
which has also failed repeatedly in
practice~\cite{DBLP:journals/corr/abs-2008-04863} and
PHANTOM~\cite{DBLP:conf/aft/SompolinskyWZ21}, for which a vulnerability has been
shown~\cite{DBLP:journals/corr/abs-1805-03870} in an early version of the
protocol.

\section{Model}
\label{sec:model}

\subsection{Avalanche platform}
\label{sec:ava-plat}

We briefly review the architecture of the Avalanche
platform~\cite{documentation}.
It consists of three separate built-in blockchains, the \emph{exchange} or
\emph{X-Chain}, the \emph{platform} or \emph{P-Chain}, and the
\emph{contract} or \emph{C-Chain}.  Additionally there are a number of
subnets.  In order to participate in the protocols and validate
transactions, a party needs to stake at least 2'000\,AVAX (about 50'000\,
USD in August 2022~\cite{coinmarketcap}).

The \emph{exchange chain} or \emph{X-Chain} secures and stores
\emph{transactions} that trade digital assets, such as the native AVAX
token. This chain implements a variant of the \emph{Avalanche consensus protocol} that
only partially orders the transactions and that is the focus of this work.
All information given here refers to the original specification of Avalanche~\cite{avalanche}.

The \emph{platform chain} or \emph{P-Chain} secures \emph{platform
  primitives}; it manages all other chains, designates parties to become
validators or removes them again from the validator list, and creates or
deletes wallets.  The P-Chain implements the \emph{Snowman} consensus
protocol: this is a special case of Avalanche consensus that always
provides total order, like traditional blockchains.  It is not explained in
the whitepaper and we do not describe it further here.

The \emph{C-Chain} hosts \emph{smart contracts} and runs transactions on an
Ethereum Virtual Machine (EVM).  It also implements the Snowman consensus
protocol of Avalanche and totally orders all transactions and blocks.

\subsection{Communication and adversary}
\label{sec:adversary}
We now abstract the Avalanche consensus protocol and consider a network of
$n$ parties $\CN = \{p, q, \dots \}$ that communicate with each other by
sending messages. An adversary may \emph{corrupt} up to $f$ of these
parties and cause them to behave \emph{maliciously} and diverge arbitrarily
from the protocol.  Non-corrupted parties are known as \emph{honest},
messages and transactions sent by them are referred to as \emph{honest}.
Analogously, corrupted parties send \emph{malicious} transactions and
messages.  The parties may access a low-level functionality for sending
messages over authenticated point-to-point links between each pair of
parties. In the protocol, this functionality is accessed by two events
\emph{send} and \emph{receive}. Parties may also access a second low-level
functionality for broadcasting messages through the network by
gossiping, accessed by the two
events \emph{gossip} and \emph{hear} in the protocol. Both primitives are subject to
network and timing assumptions.  We assume \emph{partial synchrony}, as in
the original Avalanche whitepaper~\cite{avalanche}.  Messages are
delivered according to an exponential distribution, that is, the amount of
time between the sending and the receiving of a message follows an
exponential distribution with unknown parameter to the parties.  However,
messages from corrupted parties are not affected by this delay and will be
delivered as fast as the adversary decides. This model differs from the
traditional definition of partial
synchrony~\cite{DBLP:journals/jacm/DworkLS88}, since the adversary does not
possess the ability to delay honest messages as it pleases.

\subsection{Abstractions}
\label{sec:Abstractions}

The \emph{payload transactions} of Avalanche are submitted by users and
built according to the \emph{unspent transaction output} (\emph{UTXO})
model of Bitcoin~\cite{bitcoin}. A payload transaction~\var{tx} contains a
set of \emph{inputs}, a set of \emph{outputs}, and a number of digital
signatures.  Every input refers to a position in the output of a
transaction executed earlier; this output is thereby \emph{spent} (or
\emph{consumed}) and distributed among the outputs of~\var{tx}.  The
balance of a user is given by the set of unspent outputs of all
transactions (UTXOs) executed by the user (i.e., assigned to public keys
controlled by that user).  A payload transaction is valid if it is properly
authenticated and none of the inputs that it consumes has been consumed yet
(according to the view of the party executing the validation).

Blockchain protocols are generally formalized as \emph{atomic broadcast},
since every party running the protocol outputs the same ordered list of
transactions. However, the transaction sequences output by two different
parties running Avalanche may not be exactly the same because
Avalanche allows more flexibility and does not require a total
order. Avalanche only orders transactions that causally depend on each
other. Thus, we abstract Avalanche as a \emph{generic broadcast} according
to Pedone~and~Schiper~\cite{DBLP:conf/wdag/PedoneS99}, in which the
total-order property holds only for \emph{related} transactions as follows.

\begin{definition}
 \label{def:related}
  Two payloads
   $\var{tx}$ and $\var{tx}'$ are said to be
  \emph{related}, denoted by $\var{tx}\sim\var{tx}'$, if $\var{tx}$
  consumes an output of $\var{tx}'$ or vice versa.
\end{definition}

Our generic broadcast primitive is accessed through the events
$\var{broadcast}(\var{tx})$ and $\var{deliver}(\var{tx})$. Similar to other
blockchain consensus protocols, it defines an ``external'' validity
property and introduces a predicate $V$ that determines whether a
transaction is valid~\cite{DBLP:conf/crypto/CachinKPS01}.  

\begin{definition}
  \label{def:validity}
  A payload $\var{tx}$ satisfies the validity predicate of Avalanche if all
  the cryptographic requirements are fulfilled and there is no other
  delivered payload with any input in common with $\var{tx}$.
\end{definition}

For the remainder of this work, we fix the external validation
predicate~$V$ to check the validity of payloads according to the logic of
UTXO mentioned before.

Since Avalanche is a randomized protocol, the properties of our broadcast
abstraction need to be fulfilled only with all but negligible probability.

\begin{definition}\label{def:gb}  
  A protocol solves \emph{validated generic broadcast} with validity
  predicate~$V$ and relation $\sim$ if it satisfies the following
  conditions, except with negligible probability:
  \begin{description}
  \item[Validity.] If a honest party \op{broadcasts} a payload transaction $\var{tx}$, then it eventually \op{delivers} $\var{tx}$.
  \item[Agreement.] If a honest party \op{delivers} a payload transaction $\var{tx}$, then all honest parties eventually \op{deliver}~$\var{tx}$.
  \item[Integrity.] For any payload transaction $\var{tx}$, every honest party \op{delivers} $\var{tx}$  at most once, and only if $\var{tx}$ was previously \op{broadcast} by some party.
  \item[Partial order.] If honest parties $p$ and $q$  both \op{deliver} payload transactions $\var{tx}$ and $\var{tx}'$ such that $\var{tx}\sim\var{tx}'$, then $p$ \op{delivers} $\var{tx}$ before $\var{tx}'$ if and only if $q$ \op{delivers} $\var{tx}$ before $\var{tx}'$.
  \item[External validity.] If a honest party \op{delivers} a payload transaction $\var{tx}$, then $V(\var{tx})=\true$.
  \end{description}
\end{definition}
  
Note that different instantiations of the relation $\sim$ transform the
generic broadcast primitive into well-known primitives. For instance, when
no pair of transactions are related, generic broadcast degenerates to
reliable broadcast. Whereas when every two transactions are related,
generic broadcast transforms into atomic broadcast.
In our context, broadcasting corresponds to submitting a payload
transaction to the network, whereas delivering corresponds to accepting a
payload and appending it to the ledger.

The Avalanche protocol augments payload transactions to \emph{protocol
  transactions}.  A protocol transaction additionally contains a set of
\emph{references} to previously executed protocol transactions, together
with further attributes regarding the execution.  A protocol transaction in
the implementation contains a batch of payload transactions, but this
feature of Avalanche is ignored here, since it affects only efficiency.
Throughout this paper, \emph{transaction} refers to a protocol transaction,
unless the opposite is indicated, and \emph{payload} means simply a payload
transaction.

A transaction references one or multiple previous transactions, unlike
longest-chain protocols, in which each transaction has a unique
parent~\cite{bitcoin}.  An execution of the Avalanche protocol will
therefore create a directed acyclic graph (DAG) that forms its ledger data
structure.

Given a protocol transaction $T$, all transactions that it references are
called the \emph{parents} of $T$ and denoted by $\var{parents}(T)$. The
parents of $T$ together with the parents of those, recursively, are called
the \emph{ancestors} of $T$, denoted by $\var{ancestors}(T)$. Analogously,
the transactions that have $T$ as parent are called the \emph{children} of
$T$ and are denoted by $\op{children}(T)$. Finally, the children of $T$
together with their recursive set of children are called the
\emph{descendants} of $T$, denoted by $\var{descendants}(T)$.

Note that two payload transactions $\var{tx}_1$ and $\var{tx}_2$ in
Avalanche that consume the same input are not related, unless the condition
of Definition~\ref{def:related} is fulfilled.  However, two Avalanche
payloads consuming the same output \emph{conflict}.  For each transaction
$T$, Avalanche maintains a set $\op{conflictSet}[T]$ of transactions that
conflict with~$T$.

\begin{figure}
  \centering
  \includegraphics[width=0.75\textwidth]{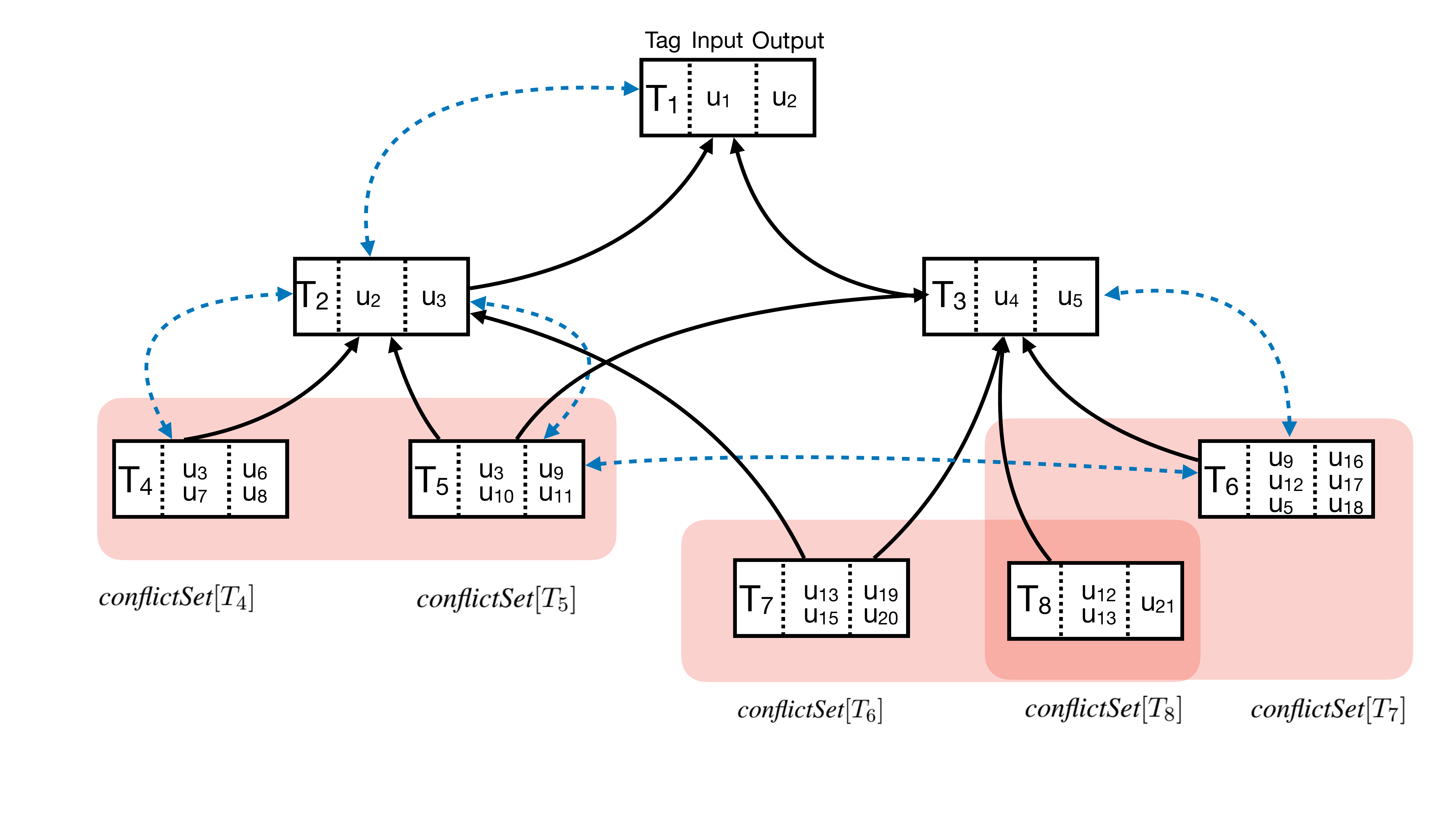}
  \vspace*{-4ex}
  \caption{The UTXO model, conflicting transactions, and related transactions in Avalanche. The eight transactions are labeled $T_1,...,T_8$. Each transaction is divided into three parts: the left part is a tag $T_i$ to identify the transaction, the middle part is its set of inputs, and the right part is its set of outputs. The solid arrows indicate the references added by the protocol, showing the parents of each transaction. For instance, $T_5$ references $T_2$ and $T_3$ and has them as parents. The dashed double-arrows indicate related transactions. For example, $T_5$ and $T_2$ are related because $u_3$ is created by $T_2$ and consumed by $T_5$. The conflict sets are denoted by the shaded (red) rectangles. As illustrated, conflict sets can be symmetric, as for $T_4$ and $T_5$, where the conflict sets are identical ($\var{conflictSet}[T_{4}]=\var{conflictSet}[T_{5}]$) or asymmetric, as for $T_6$, $T_7$, and $T_8$ where $\var{conflictSet}[T_{6}]\cup \var{conflictSet}[T_{7}]=\var{conflictSet}[T_{8}]$.}
 \label{fig:example}
\end{figure}

\section{A description of the Avalanche protocol}
\label{sec:description}

Avalanche's best-known quality is its efficiency. Permissionless
consensus protocols, such as those of Bitcoin and
Ethereum, are traditionally slow, suffer from low throughput and high latency,
and consume large amounts of energy, due to
their use of proof-of-work (PoW).  Avalanche substitutes PoW with a random
sampling mechanism that runs at network speed and that has every party
adjust its preference to that of a (perceived) majority in the
system. Avalanche also differs from more traditional blockchains by forming
a DAG of transactions instead of a chain.

\subsection{Overview}
\label{sec:details}

Avalanche is structured around its \emph{polling} mechanism. In a nutshell,
party $u$ repeatedly selects a transaction~$T$ and sends a \emph{query}
about it to $k$ randomly selected parties in the network. If a majority of
them send a positive reply, the query is successful and the transaction
contributes to the security of other transactions. Otherwise, the transaction is
still processed but does not contribute to the security of any other
transactions.  Then the party selects a new transaction and repeats the
procedure.  A bounded number of such polls may execute concurrently.
Throughout this work the terms ``poll'' and ``query'' are interchangeable.

In more detail, the protocol operates like this.  Through the \emph{gossip} functionality,
every party is aware of the network membership~\CN.  A party locally stores all those
transactions processed by the network that it knows.  The transactions form
a DAG through their references as described in the previous section.

Whenever a user submits a payload transaction \var{tx} to the network, the
user actually submits it through a party~$u$. Then, $u$ randomly selects
a number of leaf nodes from a part of the DAG known as the \emph{virtuous
  frontier}; these are the leaf nodes that are not part of any conflicting
set.  Party~$u$ then extends \var{tx} with references to the selected nodes
and thereby creates a transaction~$T$ from the payload
transaction~\var{tx}.  Next, $u$ sends a \str{Query} message with $T$ to
$k$ randomly, according to stake, chosen parties in the network and waits for their replies in
the form of \str{Vote} messages.  When a party receives a query for~$T$
and if $T$ and its ancestors are \emph{preferred}, then the party
replies with a positive vote.
The answer to this query depends exclusively on the
status of $T$ and its \emph{ancestors} according to the local view of the
party that replies.
Moreover, the definition of \emph{preferred} is
non-trivial and will be explained further below. If the polling party
receives more than
$\alpha>\frac{k}{2}$ positive votes, the poll is defined to be
successful. 

Every party $u$ running the Avalanche protocol sorts transactions of its DAG into
conflict sets.
\begin{definition}
  The \emph{conflict set} $\op{conflictSet}[T]$ of a given transaction $T$ is the set
  of transactions that have an input in common with $T$ (including $T$
  itself).
\end{definition}
Note that even if two transaction $T$ and $T'$ consume one common
transaction output and thus
conflict, their conflict sets $\op{conflictSet}[T]$ and $\var{conflictSet}[T']$ can differ, since
$T$ may consume outputs of further transactions.  (In
Figure~\ref{fig:example}, for example, $T_8$ conflicts with $T_6$ and
$T_7$, although $T_7$ conflicts with $T_8$ but not with $T_6$.)

Decisions on accepting transactions are made as follows.  For each of its
conflict sets, a party selects one transaction and designates it as
\emph{preferred}.  This designation is parametrized by a \emph{confidence value}~$d[T]$ of $T$,
 which is updated after each transaction query. If the confidence
value of some conflicting transaction $T^\ast$ surpasses $d[T]$, then
$T^\ast$ becomes the preferred transaction in the conflict set.

It has been shown~\cite{polling,avalanche} that regardless of the initial distribution of such confidence
values and preferences of transactions, this mechanism converges.  For the
transactions of one conflict set considered in isolation, this implies that all honest parties eventually prefer the same
transaction from their local conflict sets.  (The actual protocol has to
respect also dependencies among the transactions; we return to this
later.)

To illustrate this phenomenon, assume that there exist only two
transactions $T$ and $T'$ and that half of the parties prefer $T$,
whereas the other half prefers~$T'$.  This is the worst-case
scenario. Randomness in sampling breaks the tie. Without loss of
generality, assume that parties with preferred transaction~$T$ are queried
more often. Hence, more parties consider $T$ as preferred as a consequence.
Furthermore, the next time when a party samples again, the probability of
hitting a party that prefers $T$ is higher than hitting one that
prefers~$T'$.  This is the ``snowball'' effect that leads to ever
more parties preferring $T$ until every party prefers~$T$.

This preferred transaction is the candidate for acceptance and
incorporation into the ledger. The procedure is parametrized by a confidence
counter for each conflict set,
which reflects the probability that $T$ is the preferred transaction in the
local view of the party.  The party increments the confidence counter
whenever it receives a positive vote to a query on a descendant of $T$; the
counter is reset to zero whenever such a query obtains a negative vote.
When this counter overcomes a given threshold, $T$ is accepted and
its payload is added to the ledger.
We now present a detailed description of the protocol and refer to the
pseudocode in Algorithm~\ref{algo:state}--\ref{algo:aux3}.

\subsection{Data structures}

The information presented here has been taken from the whitepaper~\cite{avalanche}, the source code~\cite{source}, or the official documentation~\cite{documentation}.

\para{Notation}
We introduce the notation used in the remaining sections including the
pseudocode.  For a variable~$a$ and a set $\CS$, the notation
$a\overset{R}{\gets} \CS$ denotes sampling $a$ uniformly at random from
$\CS$.  We frequently use hashmap data structures: A hashmap associates
keys in a set \CK with values in \CV and is denoted by
$\var{HashMap}[ \CK \to \CV ]$.  For a hashmap \CF, the notation $\CF[K]$
returns the entry stored under key~$K \in \CK$; referencing an unassigned key gives
a special value~$\bot$.

We make use of timers throughout the protocol description.  Timers are
created in a stopped state.  When a timer has been started, it produces a
\op{timeout} event once after a given duration has expired and then
stops. A timer can be (re)started arbitrarily many times. Stopping a timer
is idempotent.

\para{Global parameters}
We recall that we model Avalanche as run by an immutable set of parties \CN
of size $n$. There are more three global parameters: the number $k$ of
parties queried in every poll, the majority threshold $\alpha> \frac{k}{2}$
for each poll, and the maximum number $\var{maxPoll}$ of concurrent polls.

\para{Local variables}
Queried transactions are stored in a set \CQ, the subset $\CR\subset\CQ$ is defined to be the set of \emph{repollable} transactions, a feature that is not explained in the original paper~\cite{avalanche}. The number of active polls is tracked in a variable $\var{conPoll}$. The parents of a transaction are selected from the \emph{virtuous frontier}, \CVF, defined as the set of all \emph{non-conflicting} transactions that have no known descendant and whose ancestors are preferred in their respective conflict sets. A transaction is non-conflicting if there is no transaction in the local DAG spending any of its inputs. For completeness, we recall that conflicting transactions are sorted in $\var{conflictSet[T]}$ formed by transactions that conflict with $T$, i.e., transactions which have some input in common with~$T$.

Transactions bear several attributes related to queries and transaction preference. A \emph{confidence value} $d[T]$ is defined to be the number of positive queries of $T$ and its descendants. Given a conflict set $\op{conflictSet}[T]$, the variable $\op{pref}[\op{conflictSet}[T]]$, called \emph{preferred transaction}, stores the transaction with the highest confidence value in $\op{conflictSet}[T]$. The variable $\op{last}[\op{conflictSet}[T]]$ denotes which transaction was the preferred one in $\op{conflictSet}[T]$ after the most recent update of the preferences. The preferred transaction is the candidate for acceptance in each conflict set, the acceptance is modeled by a counter $\op{cnt}[\op{conflictSet}[T]]$. Once accepted, a transaction remains the preferred one in its conflict set forever.

\subsection{Detailed description}
\label{sec:subroutines}

Each transaction does through three phases during the consensus protocol: query of transactions, reply to queries, and update of preferences. All of the previous phases call the same set of functions.

\para{Functions} 
The function $\op{updateDAG}(T)$ sorts the transactions in the
corresponding conflict sets. The function $\op{preferred}(T)$
(L~\ref{l:fun_pref}) outputs \true if $T$ is the preferred transaction in
its conflict set and \false otherwise. The function
\op{stronglyPreferred($T$)} (L~\ref{l:strong1}) outputs \true if and only
if $T$, and everyone of its ancestors is the preferred transaction in its
respective conflict set.

The function $\op{acceptable}(T)$ (L~\ref{l:isAcceptedCondition1})
determines whether $T$ can be accepted and its payload added to the ledger
or not. Transaction $T$ is considered accepted when one of the two
following conditions is fulfilled:
\begin{itemize}
\item $T$ is the unique transaction in its conflict set, all the
  transactions referenced by $\CT$ are considered accepted, and
  $\op{cnt}[\op{conflictSet}[T]]$ is greater or equal than $\beta_1$.
\item $\op{cnt}[\op{conflictSet}[T]]$ is greater or equal than $\beta_2$.
\end{itemize}
Finally, the function $\op{updateRepollable}()$ (L~\ref{l:repollable1})
updates the set of repollable transactions. A transaction $T$ is repollable
if $T$ has already been accepted; or all its ancestors are preferred, a
transaction in its conflict set has not already been accepted, and no
parent has been rejected

\para{Transaction query}
A party in Avalanche progresses only by querying transactions. In each of
these queries, party $u$ selects a random transaction $T$
(L~\ref{l:query1}), from the set of transactions that $u$ has not previously
queried by $u$. Then, it samples a random subset $\CS[T]\subset\CN$ of $k$ parties
from the set of parties running the Avalanche protocol and sends each a
\msg{query}{T} message. In the implementation of the protocol, $u$ performs
$\var{numPoll}$ simultaneous queries. The repoll functionality
(L~\ref{l:query1}--\ref{l:query2}) consists of performing several
simultaneous transactions. When $u$ does not know of any transaction that has
not been queried, $u$ queries a transaction that has not been accepted
yet. The main idea behind this functionality is to utilize the network when
this is not saturated.  The repoll functionality
(L~\ref{l:query1}--\ref{l:query2}) constitutes one of the most notable
changes from Avalanche's whitepaper~\cite{avalanche}.

\para{Query reply}
Whenever $u$ receives a query message with transaction $T$, party $u$
replies with a message \msg{vote}{u,T,\op{stronglyPreferred($T$)}}
containing the output of the binary function \op{stronglyPreferred($T$)}
according to its local view (L~\ref{l:strong1}).

\para{Update of preferences}
Party $u$ collects the replies
\msg{vote}{v,T,\op{stronglyPreferred($T$)}}, and counts the number of
positive votes. On the one hand, if the number of positive votes
overcomes the threshold $\alpha$ (L~\ref{l:positive1}), the query is considered successful. In
this case party $u$ loops over $T$ and all its ancestors~$T'$,
increasing the confidence level $d[T']$ by one. If $T'$ is the
preferred transaction in its conflict set, then party $u$ increases the
counter for transaction $\op{cnt}[\var{conflictSet}[T']]$ by
one.  Subsequently, $u$ checks whether $T'$ has also previously
been the preferred transaction in its conflict set.
And when $T'$ is not the preferred transaction according to
the most recent query, party $u$ will set the counter to one
(L~\ref{l:positive1}--\ref{l:positive2}), in order to ensure
that $\op{cnt}[\var{conflictSet}[T']]$ correctly reflects the number
of consecutive successful queries of descendants of $T'$.

On the other hand, if $u$ receives more than $k-\alpha$ negative votes,
party $u$ loops also over $T$ and its ancestors, and sets their
counters $\op{cnt}[\var{conflictSet}[T']]$ to zero as if
to indicate that $T'$ and the other transactions should not be accepted yet.
(L~\ref{l:negative1}--\ref{l:negative2}).

\para{Acceptance of transactions} Party $u$ accepts transaction $T$ when
its counter $\op{cnt}[\var{conflictSet}[T]]$ reaches a certain threshold
$\beta_1$ or $\beta_2$. If $T$ is the only transaction in its conflicting set and
all its parents have already been accepted, then $u$ accepts $T$ if
$\op{cnt}[\var{conflictSet}[T]]\geq \beta_1$, otherwise $u$ waits until the counter 
overcomes a higher value $\beta_2$.

\para{No-op transactions}
The local DAG is modified whenever a poll is finalized. In particular, only
the queried transaction ans its ancestors are modified. Avalanche makes use
of \emph{no-op transactions} to modify all the transactions in the
DAG. After finalizing a poll, party $u$ queries the network with all the
transactions in the virtuous frontier whose state has not been modified, in
a sequential manner.

\begin{algo*}[b!]
  \begin{numbertabbing}
    xxxx\=xxxx\=xxxx\=xxxx\=xxxx\=xxxx\=xxxx\kill
      \textbf{Global parameters and state}\\
      $\CN$ \` // set of parties\label{}\\
  	  $\var{maxPoll}\in\mathbb{N}$ \` // maximum number of concurrent polls, default value 4\label{}\\
  	  $k\in\mathbb{N}$  \` // number of parties queried in each poll, default value 20\label{}\\
  	  $\alpha\in\{\lceil\frac{k+1}{2}\rceil,...,k\}$ \` // majority threshold for queries, default value 15\label{}\\
  	  $\beta_1\in\mathbb{N}$ \` // threshold for early acceptance, default value 15\label{}\\
 	  $\beta_2\in\mathbb{N}$ \` // threshold for acceptance, default value 150\label{}\\
      $\CT\gets\emptyset$ \` // set of known transactions\label{}\\
      $\CQ\subset\CT\gets\emptyset$ \` // set of queried transactions\label{}\\
      $\CR\subset\CQ\gets\emptyset$ \` // set of repollable transactions\label{}\\
      $\CD\subset\CT\gets\emptyset$ \` // set of no-op transactions to be queried\label{}\\
      $\CVF\subset\CQ\gets\emptyset$ \` // set of transactions in the virtuous frontier\label{}\\
      $\var{conPoll}\in\mathbb{N}\gets 0$  \` // number of concurrent polls performed\label{}\\
      $\var{conflictSet}:\op{HashMap}[\CT\to 2^{\CT}]$ \` // conflict set\label{}\\
      $\CS:\op{HashMap}[\CT\to\CN]$ \` // set of sampled parties to be queried with a transaction\label{}\\
      $\var{votes}:\op{HashMap}[\CT\times\CN\to \{\false,\true\} ]$ \` // variable to store the replies of queries\label{}\\

      $d:\op{HashMap}[\CT\to\mathbb{N}]$ \` // confidence value of a transaction\label{}\\
      $\var{pref}:\op{HashMap}[2^\CT\to\CT]$ \` // preferred transaction in the conflict set\label{}\\
      $\var{last}:\op{HashMap}[2^\CT\to\CT]$ \` // preferred transaction in the last query\label{}\\
      $\var{cnt}:\op{HashMap}[2^\CT\to\mathbb{N}]$ \` // counter for acceptance of the conflict set\label{}\\ 
      $\var{accepted}:\op{HashMap}[\CT\to\{\false,\true\}]$  \` // indicator that a transaction is accepted\label{}\\
      $\var{timer}:\op{HashMap}[\CT\to\{\var{timers}\}]$ \` // timer for the query of transactions\label{}
    \end{numbertabbing}
    \caption{Avalanche (party $u$), state}
    \label{algo:state}
  \end{algo*}

\begin{algo*}
  \begin{numbertabbing}
    xxxx\=xxxx\=xxxx\=xxxx\=xxxx\=xxxx\=xxxx\kill
      \textbf{upon} $\op{broadcast}(\var{tx})$ \textbf{do}\label{l:submit1}\\
      \> \textbf{if} $V(\var{tx})$ \textbf{then}\label{}\\
      \>\> $T\gets(\var{tx},\CVF)$ \` // up to a maximum number of parents\label{}\\
      \>\> $\CT\gets\CT\cup\{T\}$\label{}\\
      \>\> $\var{accepted}[T]\gets\false$\label{}\\
      \>\>$\op{updateDAG}(T)$\label{}\\ 
      \>\> gossip message \msg{broadcast}{T}\label{l:submit2}\\
      \\     
      \textbf{upon} hearing message \msg{broadcast}{T} \textbf{do}\label{l:hear1}\\
      \>\textbf{if} $T\not\in\CT$ \textbf{do}\label{l:hear}\\
      \>\>$\CT\gets \CT\cup\{T\}$\label{}\\
      \>\> $\var{accepted}[T]\gets\false$\label{l:hear2}\\ 

      \\
      \textbf{upon} $\var{conPoll}<\var{maxPoll}$ \textbf{do}\label{l:query1}\\
      \> $\var{conPoll} \gets \var{conPoll}+1$\label{}\\
      \>\textbf{if} $\CD \neq \emptyset$ \textbf{then}
      \` // prefer no-op transactions\label{l:no-op}\\
      \>\> $T \gets \;\text{least recent transaction in}\ \CD$\label{}\\
      \>\textbf{else if} $\CT\setminus\CQ \neq \emptyset$ \textbf{then}
      \` // take any not yet queried transaction\label{}\\
      \>\> $T\overset{R}{\gets}\CT\setminus\CQ$\label{}\\
      \>\> $\op{d}[T]\gets 0$\label{}\\ 
      \>\textbf{else}
      \` // all transaction queried already, take one of them\label{l:repol1}\\
      \>\> $\op{updateRepollable}()$\label{}\\
      \>\> $T\overset{R}{\gets}\CR$\label{l:repol2}\\
      \>$\CS[T] \gets\op{sample}(\CN\setminus\{u\}, k)$\` // sample $k$ parties randomly according to stake\label{}\\
      \> \op{send} message \msg{Query}{T} to all parties $v\in \CS[T]$\label{l:sendTransaction}\\
      \>$\CD\gets\CD\cup\{(\perp ,\CVF\setminus\{T\})\}$\` // create a no-op transaction\label{}\\
      \> \textbf{start} $\op{timer}[T]$\` // duration $\Delta_{\var{query}}$\label{}\\
      \>$\CQ\gets\CQ\cup \{T\}$\label{}\\
      \>$\op{updateDAG}(T)$\label{l:query2}\\
      \\
      \textbf{upon} receiving message \msg{Query}{T} from party $v$ \textbf{do}\label{l:receiveTransaction}\\
      \>\op{send} message \msg{Vote}{u,T,\op{stronglyPreferred($T$)}} to party $v$\label{}\\
      \\
       \textbf{upon} receiving message \msg{Vote}{v,T,w} such that $v\in\CS[T]$ \textbf{do}\` // $w$ is the vote\label{l:receiveVote}\\
       \>$\op{votes}[T,v]\gets w$\` // $w\in\{\false,\true\}$\label{}
           \end{numbertabbing}
    \caption{Avalanche (party $u$), part 1}
    
   \label{algo:aux1}
  \end{algo*}

    \begin{algo*} 
    \begin{numbertabbing}
      xxxx\=xxxx\=xxxx\=xxxx\=xxxx\=xxxx\=xxxx\kill
      \textbf{upon} $\exists T \in \CT$ such that $\bigl| \{v \in \CS[T] \mid \op{votes}[T,v] = \true\} \bigr| \geq \alpha$ \textbf{do}\` // query of $T$ is successful\label{l:positive1}\\
      \> \textbf{stop} $\op{timer}[T]$\label{}\\
      \> $\op{votes}[T, *] \gets \bot$\` // remove all entries in \op{votes} for $T$\label{}\\ 
      \> $\var{S}[T]\gets [\ ]$\` // reset $S$ for $T$\label{}\\ 
      \>$d[T]\gets d[T]+1$\label{}\\
      \> \textbf{for} $T'\in\var{ancestors}(T)$ \textbf{do}\` // all ancestors of $T$\label{}\\
      \>\> $d[T']\gets d[T']+1$\label{}\\
      \>\>\textbf{if} $d[T']>d[\op{pref}[\var{conflictSet}[T']]]$ \textbf{then}\label{}\\
      \>\>\> $\op{pref}[\var{conflictSet}[T']]\gets T'$\label{}\\
      \>\>\textbf{if} $T'\neq \op{last}[\var{conflictSet}[T']]$ \textbf{then}\label{l:change}\\
      \>\>\> $\op{last}[\var{conflictSet}[T']]\gets T'$\label{}\\
      \>\>\>$\op{cnt}[\var{conflictSet}[T']]\gets1$\label{}\\
      \>\>\textbf{else}\label{}\\
      \>\>\> $\op{cnt}[\var{conflictSet}[T']]\gets \op{cnt}[\var{conflictSet}[T']]+1$\label{}\\
      \> $\var{conPoll}\gets \var{conPoll}-1$\label{l:positive2}\\
      \\
      \textbf{upon} $\exists T \in \CT$ such that $\bigl| \{v \in \CS[T] \mid \op{votes}[T,v] = \false \} \bigr| > k - \alpha$ \textbf{do}\` // query of $T$ failed\label{l:negative1}\\
      \> \textbf{stop} $\op{timer}[T]$\label{}\\
      \> $\op{votes}[T, *] \gets \bot$\` // remove all entries in \op{votes} for $T$\label{}\\ 
      \> $\var{S}[T]\gets [\ ]$\` // reset $S$ for $T$\label{}\\ 
      
      \> \textbf{for} $T'\in\var{ancestors}(T)$ \textbf{do}\` // all ancestors of $T$\label{}\\
      \>\>$\op{cnt}[\var{conflictSet}[T']]\gets 0$\label{l:negative2}\\
      \\
      \textbf{upon} $\exists T\in\CT$ such that $\op{acceptable(T)}\land \neg\var{accepted}[T]$ \textbf{do}\` // $T$ can be accepted\label{l:isAccepted1}\\
      \> $(\var{tx},\var{parents})\gets T$\label{}\\
      \>\textbf{if} $V(\var{tx})$ \textbf{then}\label{}\\
      \>\> $\var{accepted}[T]\gets\true$\label{}\\
      \>\>\op{deliver} $\var{tx}$\label{l:isAccepted2}\\
      \\ 
      \textbf{upon} timeout from $\op{timer}[T]$ \textbf{do}\label{} \` // not enough votes on $T$ received\label{l:timeout1}\\
      \>$\CQ\gets\CQ\setminus\{T\}$\label{}\\
      \> $\op{votes}[T, *] \gets \bot$\` // remove all entries in \op{votes} for $T$\label{}\\ 
      \> $\var{S}[T]\gets [\ ]$\` // do not consider more votes from this query\label{l:timeout2}\\
             \end{numbertabbing}
    \caption{Avalanche (party $u$), part 2}
    
   \label{algo:aux2}
  \end{algo*}

    \begin{algo*} 
    \begin{numbertabbing}
        xxxx\=xxxx\=xxxx\=xxxx\=xxxx\=xxxx\=xxxx\kill    
      ~\textbf{function} $\op{updateDAG}(T)$\label{}\\
      \> $\CVF\gets\text{set of non-conflicting leaves in the DAG}$\label{}\\ 
      \> $\op{conflictSet}[T]\gets \emptyset$\label{}\\
      \> \textbf{for} $\ T'\in\CT$ such that $T'\neq T$ and $T'$ has a common input with  $T$ \textbf{do}\label{}\\ 
      \>\>$\op{conflictSet}[T]\gets\op{conflictSet}[T]\cup \{T'\}$\label{}\\
      \>\> $\var{conflictSet}[T']\gets\var{conflictSet}[T']\cup \{T\}$\label{}\\
      \> \textbf{if} $\op{conflictSet}[T] = \emptyset$ \textbf{then}\` // $T$ is non-conflicting\label{}\\ 
      \>\> $\op{pref}[\op{conflictSet}[T]]\gets T$\label{}\\
      \>\> $\op{last}[\op{conflictSet}[T]]\gets T$\label{}\\
      \>\> $\op{cnt}[\op{conflictSet}[T]]\gets 0$\label{}\\ 
      \> $\op{conflictSet}[T]\gets\op{conflictSet}[T]\cup\{T\}$\label{}\\ 
      \\

      ~\textbf{function} $\op{getParents}(T)$\label{l:get}\\
      \> $(\var{tx},\var{parents})\gets T$\label{}\\
      \>\textbf{return} $\var{parents}$\` // set of parents stored in $T$ \label{}\\ 
      \\

      ~\textbf{function} $\op{preferred}(T)$\label{l:fun_pref}\\
      \> \textbf{return} $T \stackrel{?}{=} \op{pref}[\op{conflictSet}[T]]$ \label{}\\
      \\
      ~\textbf{function} $\op{stronglyPreferred}(T)$\label{l:strong1}\\
      \> \textbf{return} $\underset{T'\in\var{ancestors}(T)}{\bigwedge} \op{preferred}(T')$\label{}\\
      \\
      ~\textbf{function} $\op{acceptable}(T)$\label{l:isAcceptedCondition1}\\
      \>\textbf{return} $\bigl( \bigl| \op{conflictSet}[T] \bigr| = 1
        \land \op{cnt}\bigl[\op{conflictSet}[T]\bigr] \geq \beta_1 \bigr)
        \land {\underset{T'\in\ \var{parents}(T)}{\bigwedge}}
          \op{acceptable}(T')$ \label{}\\
      \>\>$\mbox{} \lor \op{cnt}\bigl[\op{conflictSet}[T]\bigr]\geq\beta_2$
      \label{l:isAcceptedCondition2}\\
      \\
      ~\textbf{function} $\op{isRejected}(T)$\label{}\\
      \>\textbf{return} $\exists T'\in\CT$ such that $\forall T' \in \op{conflictSet}[T] \setminus \{T\}: \op{acceptable}(T')$ \\
      \\

      ~\textbf{function} $\op{updateRepollable}()$\label{l:repollable1}\\
      \>$\CR\gets\emptyset$\label{}\\
      \>\textbf{for} $T\in\CT$ \textbf{do}\label{}\\
      \>\>\textbf{if} $ \op{acceptable}(T) \lor \underset{T'\in\ \var{parents}(T)}{\bigwedge}\op{stronglyPreferred}(T')\land \neg\op{isRejected}(T')$ \textbf{then}\label{}\\
      \>\>\> $\CR\gets\CR\cup\{T\}$\label{l:repollable2} 
    \end{numbertabbing}
    \caption{Avalanche, auxiliary functions}
    \label{algo:aux3}
\end{algo*}

\subsection{Life of a transaction}
\label{sec:life}

We follow an honest transaction $T$ through the protocol.
The user submits the payload transaction $\var{tx}$ to some party $u$, then $u$ adds references $\var{refs}$ to the payload transaction, creating a transaction $T=(\var{tx},\var{refs})$. These references point to  transactions in the virtuous frontier \CVF. Transaction~$T$ is then \emph{gossiped} through the network and added to the set of known transactions $\CT$(L~\ref{l:submit1}--\ref{l:submit2}). 
Party $u$  may also \emph{hear} about  new transactions through this gossip functionality. Whenever this is the case, $u$ add the transaction to its set of known transactions $\CT$ (L~\ref{l:hear1}--\ref{l:hear2}).

Party $u$ eventually selects $T$ to be processed. When this happens, $u$ \op{samples} $k$ random parties from the network and stores them in $\CS[T]$. Party $u$ \op{queries} parties in $\CS[T]$ with $T$ and starts a timer $\op{timeout}[T]$. $T$ is added to $\CQ$ (L~\ref{l:query1}--\ref{l:query2}).

Parties queried with $T$ reply with the value of the function $\op{stronglyPreferred}(T)$ (L~\ref{l:strong1}). This function answers positively (\true) if $T$ is \emph{strongly preferred}, i.e., if $T$ and all of its ancestors are the preferred transaction inside each respective conflict set. A negative answer (\false) is returned if either $T$ or any of its ancestors fail to satisfy these conditions.

Party $u$ then stores the answer from party $v$ to the query in the variable $\op{votes}[T][v]$.

\begin{itemize}
  \item If $u$ receives more than $\alpha$ positive votes, $u$ runs over all the ancestors of $T$. If the ancestor $T'$ was the most recent (or ``last'') preferred transaction in its conflict set, its counter is increased by one. Otherwise, $T'$ becomes the most recent preferred transaction and its counter is reset to one (L~\ref{l:positive1}--\ref{l:positive2}).
  \item If $u$ receives at least $k-\alpha$ \false votes, $u$ resets the counter for acceptance of all its ancestors $\op{cnt}[T']\gets 0$ (L~\ref{l:negative1}--\ref{l:negative2}).
  \item If timer $\op{timeout}[T]$ is triggered before the query is completed, the query is aborted instead. The votes are reset and every party is removed from the set $\CS[T]$, so no later reply can be considered (L~\ref{l:timeout1}--\ref{l:timeout2}).
\end{itemize}

In parallel to the previous procedure, party $u$ may perform up to $\var{conPoll}$ concurrent queries of different transactions.

Once $T$ has been queried, it awaits in the local view of party $u$ to be accepted. Since by assumption $T$ is honest, $\op{conflictSet}[T]=\{T\}$. Hence $T$ is accepted when $\op{cnt}\big[\op{conflictSet}[T]\big]$ reaches $\beta_1$, if all the ancestors of $T$ are already accepted, or $\beta_2$ otherwise (L~\ref{l:isAcceptedCondition1}--\ref{l:isAcceptedCondition2}). We recall that $\op{cnt}[\op{conflictSet}[T]]$ is incremented whenever a query involving a descendant of $T$ is successful. However, when a non-descendant of $T$ is queried, it may trigger a no-op transaction (L~\ref{l:no-op}) that is a descendant of $T$.

If there is no new transaction waiting to be queried, i.e., $\CT\setminus\CQ$ is empty, the party proceeds with a \emph{repollable} transaction (L~\ref{l:repol1}--\ref{l:repol2}). A \emph{repollable} transaction is one that has not been previously accepted but it is a candidate to be accepted (L~\ref{l:repollable1}--\ref{l:repollable2}).

\section{Security analysis}
\label{sec:analysis}

Avalanche deviates from the established PoW protocols and uses a different
structure.  Its security guarantees must be assessed differently.  The
bedrock of security for Avalanche is random sampling.

\subsection{From Snowball to Avalanche}

The Avalanche protocol family includes Slush, Snowflake, and
Snowball~\cite{avalanche} that implement single-decision Byzantine
consensus.  Every party \emph{proposes} a value and every party must
eventually \emph{decide} the same value for an instance.  The Avalanche
protocol itself provides a ``payment system''~\cite[Sec.~V]{avalanche}; we
model it here as generic broadcast.

The whitepaper~\cite{avalanche} meticulously analyzes the three consensus
protocols.  It shows that as long as $f = O(\sqrt{n})$, the consensus
protocols are live and safe~\cite{avalanche} based on the analysis
of random sampling~\cite{ross1996stochastic}.  On the
other hand, an adversary controlling more than $\Theta(\sqrt{n})$ parties
may have the ability to keep the network in a bivalent state.

However, the Avalanche protocol itself is introduced without a rigorous
analysis.  The most precise statement about it is that \emph{``it is easy
  to see that, at worst, Avalanche will degenerate into separate instances
  of Snowball, and thus provide the same liveness guarantee for virtuous
  transactions''}~\cite[p.~9]{avalanche}.  In fact, it is easy to see that
this is \emph{wrong} because every vote on a transaction in Avalanche is
linked to the vote on its ancestors.  The vote on a descendant $T'$
of $T$ depends on the state of~$T$.

We address this situation here through the description in the previous section
and by giving a formal description of Snowball
in Appendix~\ref{sec:snowball}.  We 
notice that one can isolate single executions of Snowball that occur inside
Avalanche as follows.  Consider an execution of Avalanche and a
transaction $T$ and define an \emph{equivalent} execution of Snowball as
the execution in which every party $u$ proposes the value $1$ if $T$  is 
preferred in their local view, proposes 0 if another transaction is,
and does not propose otherwise. 
Every party also selects the same parties in each round of snowball and for a query
with $T$, for a query with a transaction that conflicts with $T$, or for
any query with a descendant of these two.

\begin{lemma}
  \label{counters}
  If party $u$ delivers an honest transaction in Avalanche, then $u$
  decided 1 in the equivalent execution of Snowball with threshold
  $\beta_1$. Furthermore, $u$ delivers a conflicting transaction in
  Avalanche, then $u$ decides 1 in Snowball with threshold $\beta_2$.
\end{lemma}

\begin{proof}
  By construction of the Avalanche and Snowball protocols in the whitepaper~\cite{avalanche},
  the counter for acceptance of value 1 in Snowball is always greater or equal than
  the counter for acceptance in Avalanche. Since a successful query in
  Avalanche implies a successful query in Snowball, if an honest
  transaction in Avalanche is delivered, the counter in the equivalent
  Snowball instance is at least $\beta_1$.  Analogously, if a conflicting
  transaction in Avalanche is delivered, then the counter in Snowball is at
  least $\beta_2$. Hence, a party in Snowball would decide 1 with the
  respective thresholds.
\end{proof}

Looking ahead, we will introduce a modification of Avalanche that ensures the
complete equivalence between Snowball and Avalanche.  We first assert
some safety properties of the Avalanche protocol.

\begin{theorem}
 \label{theo:reduction}
 Avalanche satisfies integrity, partial order, and external validity of a
 generic broadcast for payload transactions under relation $\sim$ and
 UTXO-validity.
\end{theorem}

\begin{proof}
  The proof is structured by property:
  \begin{description}
  \item[Integrity.] We show that every payload is delivered at most once.
    A payload \var{tx} may potentially be delivered multiple times in two
    ways: different protocol transactions that both carry \var{tx} may be
    accepted or \var{tx} is delivered multiple times as payload of the same
    protocol transaction.

    First, we consider the possibility of accepting two different
    transactions $T_1$ and $T_2$ carrying \var{tx}.  Assume that party $u$
    accepts transaction $T_1$ and party $v$ accepts transaction $T_2$. By
    definition, $T_1$ and $T_2$ are \emph{conflicting} because they spend
    the same inputs. Using Lemma~\ref{counters}, party $u$ and $v$ decide
    differently in the equivalent execution in Snowball, which contradicts
    agreement property of the Snowball consensus~\cite{avalanche}.

    The second option is that one protocol transaction $T$ that contains
    \var{tx} is accepted multiple times. However, this is not possible
    either because \var{tx} is delivered only if
    $\var{accepted}[T] = \false$; variable $\var{accepted}[T]$ is set to
    \true when transaction $T$ is accepted
    (L~\ref{l:isAccepted1}--\ref{l:isAccepted2}).

  \item[Partial order.] Avalanches satisfies partial order because no
    payload is valid unless all payloads creating its inputs have been
    delivered (L~\ref{l:isAccepted1}--\ref{l:isAccepted2}). Transactions
    $T$ and $T'$ are related according to
    Definition~\ref{def:related} if and only if $T$ has as input (i.e.,
    spends) at least one output of $T'$, or vice versa.  This implies
    that related transactions are delivered in the same order for any
    party.
  
  \item[External validity.] The external validity property follows from
    L~\ref{l:isAccepted1}, as a payload transaction can only be delivered
    if it is valid, i.e., its inputs have not been previously spent and the
    cryptographic requirements are satisfied.
\end{description}
\end{proof}

Theorem~\ref{theo:reduction} shows that Avalanche satisfies the safety properties
of a generic broadcast in the presence of an adversary controlling $O(\sqrt{n})$
parties. A hypothetical adversary controlling more than $\Omega(\sqrt{n})$ parties 
could violate safety. It is not completely obvious how an adversary could achieve that.
Such an adversary would broadcast two conflicting transactions $T_1$ and $T_2$.
As we already discussed, and also explained in the whitepaper of Avalanche~\cite{avalanche},
such an adversary can keep the network in a bivalent state, so the adversary keeps the network
divided into two parts: parties in part $\CP_1$ consider $T_1$  preferred, and parties in part $\CP_2$
prefer $T_2$. The adversary behaves as preferring $T_1$ when communicating with parties is $\CP_1$
and as preferring $T_2$ when communicating with parties in $\CP_2$. Eventually, a party  $u\in\CP_1$ will query only parties in $\CP_1$ or the adversary for
$\beta_2$ queries in a row. Thus, $u$ will accept transaction $T_1$. Similarly, a party $v\in\CP_2$
will eventually accept transaction $T_2$. Party $u$ will \emph{deliver} the payload contained in $T_1$ and
$v$ the payload contained in $T_2$, hence violating agreement.

An adversary controlling at most $O(\sqrt{n})$ can also violate agreement, but the required behavior is more sophisticated, as we explain next.

\subsection{Delaying transaction acceptance}
\label{sec:behavior}

An adversary aims to prevent that a party~$u$ accepts an honest
transaction~$T$.  A necessary precondition for this is
$\op{cnt}[\op{conflictSet}[T]] \geq \beta_1$.  Note that whenever a
descendant of $T$ is queried, $\op{cnt}[\op{conflictSet}[T]]$ is
modified. If the query is successful (L~\ref{l:positive1}), then
$\op{cnt}[\op{conflictSet}[T]]$ is incremented by one. If the query is
unsuccessful, $\op{cnt}[\op{conflictSet}[T]]$ is reset to zero. Remark, however,
$\op{cnt}[\op{conflictSet}[T]]$ cannot be reset to one as a result of
another transaction becoming the preferred in $\op{conflictSet}[T]$
(L~\ref{l:change}) because $T$ is honest, as there exist no
transaction conflicting with $T$.

Our adversary thus proceeds by sending to $u$ a series of cleverly
generated transactions that reference~$T$. We describe these steps that
will delay the acceptance of $T$.
\begin{enumerate}
\item \textbf{Preparation phase.} The adversary submits conflicting
  transactions $T_1$ and $T_2$. For simplicity, we assume that she submits
  first $T_1$ and then $T_2$, so the preferred transaction in both conflict
  sets will be~$T_1$.  The adversary then waits until the target
  transaction $T$ is submitted.
\item \textbf{Main phase.} The adversary repeatedly sends malicious
  transactions referencing the target $T$ and $T_2$ to $u$. These
  transactions are valid but they reference a particular set of
  transactions.
\item \textbf{Searching phase.} Concurrently to the main phase, the
  adversary looks for transactions containing the same payload as $T$. If
  some are found, she references them as well from the newly generated
  transactions.
\end{enumerate}

For simplicity, we assume that the adversary knows the acceptance counter of $T$ at $u$, so she can send a malicious transaction whenever $T$ is close to being accepted.  In practice, she can guess this only with a certain probability, which will degrade the success rate of the attack. We also assume that the query of an honest transaction is always successful, which is the worst case for the adversary.

\begin{algo*}
  \begin{numbertabbing}
    xxxx\=xxxx\=xxxx\=xxxx\=xxxx\=xxxx\=xxxx\kill
    \textbf{Initialization} \\
    \> create two conflicting transactions $T_1$ and $T_2$ \label{}\\
    \> gossip two messages \msg{broadcast}{T_1} and \msg{broadcast}{T_2}\label{}\\
    \>$\CA\gets \emptyset$\label{}\\
    \\[-1ex]
    \textbf{upon} hearing message \msg{broadcast}{T} \textbf{do}\`// target transaction\label{}\\
    \>$\CA\gets\{T\}$\label{}\\
    \\[-1ex]
    \textbf{upon} $\op{cnt}[\op{conflictSet}[T]]=\lfloor\frac{\beta_1}{2}\rfloor$ in the local view of $u$ \textbf{do}\label{}\\
    \> create $\hat{T}$ such that
    $T_2 \in\var{ancestors}(\hat{T})$ and for all $T'\in\CA$,
    also $T' \in \var{ancestors}(\hat{T})$\label{}\\
    \>\op{send} message $\msg{broadcast}{\hat{T}}$ to party $u$\`// pretend to gossip the message\label{}\\\\
    \\[-1ex]
    \textbf{upon} hearing message \msg{broadcast}{\tilde{T}}
    such $\tilde{T}$ and $T$ contain the same payload \textbf{do}\label{}\\
    \>$\CA\gets\CA\cup\{\tilde{T}\}$\label{}
  \end{numbertabbing}
  \caption{Liveness attack: Delaying transaction~$T$}
  \label{algo:liveness}
\end{algo*}

After $u$ submits $T$, the adversary starts the main phase of the
attack. If $u$ queries an honest transaction~$\hat{T}$, and if $\hat{T}$
references a descendant of $T$, then $\var{cnt}[\var{conflictSet}[T]]$
increases by one.  If it does not, then $\hat{T}$ may cause $u$ to submit a
no-op transaction referencing a descendant of $T$.  Hence, honest
transactions always increase $\var{cnt}[\var{conflictSet}[T]]$ by one, this
is the worst case for an adversary aiming to delay the acceptance of $T$.

If $u$ queries a malicious transaction $\hat{T}$, honest parties reply
with their value of $\op{stronglyPreferred}(\hat{T})$. Since $T_2$ is an
ancestor of $\hat{T}$ and not the preferred transaction in its conflict set
(as we have assumed that $T_1$ is preferred), all queried parties return
$\false$. Thus, $u$ sets acceptance counter of every ancestor of
$\hat{T}$ to zero (L~\ref{l:negative1}), in particular,
$\var{cnt}[\var{conflictSet}[T]] \gets 0$.  However, since $\hat{T}$ does not
reference the virtuous frontier, $u$ submits a no-op transaction that
references a descendant of $T$, thus increasing
$\var{cnt}[\var{conflictSet}[T]]$ to one.

We show that when the number of transactions is low, in particular when
$|\CT \setminus \CQ| \leq 1$ for every party, then Avalanche may lose liveness.

\begin{theorem}
  Avalanche does not satisfy validity nor agreement of generic broadcast
  with relation $\sim$ with one single malicious party if
  $|\CT \setminus \CQ| \leq 1$ for every party.
\end{theorem}

\begin{proof}
  We consider again the adversary described above that targets $T$ and $u$.
\begin{itemize}
\item\textbf{Validity.}  Whenever $\op{cnt}[\op{conflictSet}[T]]$ in the local
  view of $u$ reaches $\lfloor\frac{\beta_1}{2}\rfloor$, the adversary
  sends a malicious transaction to party $u$, who immediately queries it
  (since $|\CT \setminus \CQ| \leq 1$). It follows that $u$ sets
  $\op{cnt}[\op{conflictSet}[T]]$ to zero and increases it intermediately
  afterwards, due to a no-op transaction. This process repeats indefinitely
  over time and prevents $u$ from delivering the payload in~$T$.

\item\textbf{Agreement.} Assume that an honest party broadcasts the payload
  contained in $T$. The adversary forces a violation of agreement by
  finding honest parties $u$ and $v$ such that
  $\op{cnt}[\op{conflictSet}[T]]=\beta_1-1$ at $v$ and
  $\op{cnt}[\op{conflictSet}[T]]<\beta_1-1$ at $u$ (such parties exist
  because in the absence of an adversary, as $\op{cnt}[\op{conflictSet}[T]]$
  increases monotonically over time). The adversary then sends an honest
  transaction~$T_h$ that references $T$ to $v$ and a malicious transaction
  $T_m$,  as described before, to~$u$.
  On the one hand, party $v$ queries $T_h$, increments
  $\op{cnt}[\op{conflictSet}[T]]$ to $\beta_1$, accepts transaction $T$,
  and delivers the payload. On the other hand, party $u$ queries $T_m$ and
  sets $\op{cnt}[\op{conflictSet}[T]]$ to one. After that, the adversary
  behaves as discussed before. Notice that $v$ has delivered the payload
  within $T$ but $u$ will never do so.
\end{itemize}
\end{proof}
An adversary may thus cause Avalanche to violate validity and
agreement. For this attack, however, the number of transactions in
the network must be low, in particular, $|\CT \setminus \CQ| \leq 1$.
In July 2022, the Avalanche network processed 
an average of 647238 transactions per day\footnote{\url{https://subnets.avax.network/stats/network}}.
Assuming two seconds per query, four times the value observed in our local implementation,
the recommended values of 30 transactions per batch, and four concurrent polls,
the condition $|\CT \setminus \CQ| \leq 1$ is satisfied $88\%$ of the time.

However, the adversary still needs to know the value of the counter for acceptance
of the different parties.
\subsection{A more general attack}
\label{sec:shortliveness}

We may relax the assumption of knowing the acceptance counters and also
send the malicious transaction to more parties through gossip.
After selecting a target transaction, the adversary continuously gossips malicious
transactions to the network instead of sending them only to one party.
For analyzing the performance of this attack, our figure of merit will be
the number of transactions to be queried by an honest party (not counting
no-ops) for confirming the target transaction~$T$.  The larger this number
becomes, the longer it will take the party until it may accept~$T$.  We
assume that $\CT\setminus\CQ\neq\emptyset$ and that a fraction $\gamma$ of
those transactions are malicious at any point in time\footnote{Avalanche
  may impose a transaction fee for processing transactions. However, since
  the malicious transactions cannot be delivered, this mechanism does not
  prevent the adversary from submitting a large number of transactions.}. A
non-obvious implication is that the repoll function never queries the same
transaction twice.

\begin{lemma}
  \label{lemma:honest}
  Avalanche requires every party to query at least $\beta_1$ transactions
  before accepting transaction~$T$ in the absence of an adversary.
\end{lemma}
\begin{proof}
The absence of an adversary carries several simplifications. Firstly, there are no conflicting transactions, thus every transaction is the preferred one in its respecting conflict set and every query is successful. Secondly, due to the no-op transactions, the counter for acceptance of every transaction in the DAG is incremented by one after each query. Finally, a transaction $T$ is accepted when its counter for acceptance reaches $\beta_1$, since the counter of the parent of any transaction reaches $\beta_1$ strictly before $T$ (L~\ref{l:isAcceptedCondition1}).
\end{proof}

\begin{lemma}
  \label{lemma:malicious}
  The average number of queried transactions before accepting
  transaction $T$ in the presence of the adversary, as described in
  the text, is at least
  \begin{equation*}
    \beta_1+\frac{1+(2+\beta_1\gamma)(1-\gamma)^{\beta_1}-(1-\gamma)^{2\beta_1}(1+\beta_1\gamma)}{\gamma(1-\gamma)^{\beta_1}(1-(1-\gamma)^{\beta_1})}.
  \end{equation*}
\end{lemma}

\begin{figure}[H]
  \centering
  \includegraphics[width=0.9\textwidth]{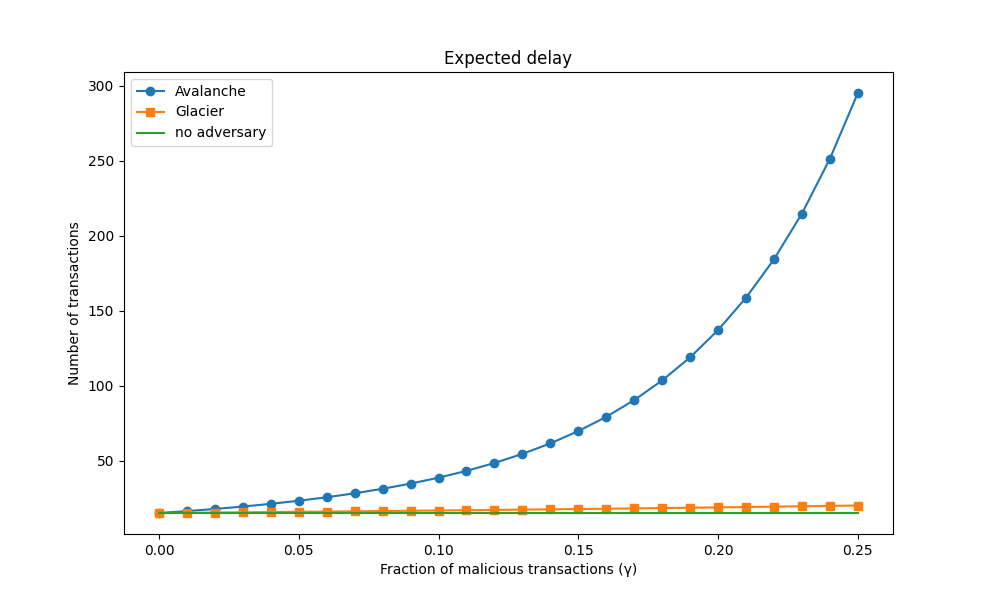}
  \vspace*{-2ex}
  \caption{Expected delay in number of transactions needed to confirm a
    given transaction with acceptance threshold $\beta_1=15$, the
    recommended value~\cite{documentation}, and assuming that the queries
    of honest transactions are successful. The (green) horizontal line
    shows $\beta_1$, the expected delay without attacker. The (blue) dotted
    line represents the expected confirmation delay in Avalanche depending
    on the fraction of malicious transactions. The (orange) squared line
    denotes the delay in Glacier (Section~\ref{app:glacier}).  }
  \label{fig:glacier}
\end{figure}

\begin{proof}
  We recall that in the worst-case scenario for the adversary, the query of
  an honest transaction increments the counter for acceptance of the target
  transaction $T$ by one, while the query of a malicious transaction,
  effectively, resets the counter for acceptance to one, as a result of a
  no-op transaction.

  Let a random variable $W$ denote the number of transactions queried by
  $u$ until $T$ is accepted, and let $X \in \{0,1\}$ model the outcome of the following
  experiment.  Party $u$ samples transactions until it picks a malicious
  transaction or until it has sampled $\beta_1-1$ honest transactions.  In
  the first case, $X$ takes the value zero, and otherwise, $X$ takes the
  value one. By definition, $X$ is a Bernoulli variable with parameter
  $p=(1-\gamma)^{(\beta_1-1)}$. Thus, the number of attempts until $X$
  returns one is a random variable~$Y$ with geometric distribution,
  $Y\sim\mathcal{G}(p)$, with the same parameter~$p$. We let
  $W_a$ be the random variable denoting the number of queried transactions
  per attempt of this experiment.  The expected number of failed attempts
  is $\E[Y]=\frac{1}{(1-\gamma)^{\beta_1}}$.  Furthermore, the probability
  that an attempt fails after sampling exactly $k$
  transactions, for $k\leq\beta_1$, is
  \begin{equation*}
    \P[W_a=k|X=0]=\frac{\gamma (1-\gamma)^{k-1}}{1-(1-\gamma)^{\beta_1}}.
  \end{equation*}
  Thus, the expected number of transactions per failed attempt can be expressed as
  \begin{equation}
    \label{eq:nfail}
    \E[W|X=0]=\frac{1-(1-\gamma)^\beta_1(1+\beta_1\gamma)}{\gamma(1-(1-\gamma)^{\beta_1})}.
  \end{equation}
  The expected number of transaction queried during a successful attempt is
  at least $\beta_1$ by Lemma~\ref{lemma:honest}.  Finally, the total expected
  number of queried transactions can be written as the expected number of
  transaction per failed attempt multiplied by the expected number of
  failed attempts plus the expected number of transactions in the
  successful attempt,
  \begin{equation}
    \label{eq:almost}
    \E[W]=\E[W_a|X=0]\cdot(\E[Y]-1)+\E[W_a|X=1]\cdot1.
  \end{equation}
  From equations~(\ref{eq:nfail})~and~(\ref{eq:almost}) and basic algebra,
  we obtain
  \begin{align*}
    \E[W]%&=\beta_1+\frac{1}{\gamma(1-(1-\gamma)^{\beta_1})(1-\gamma)^{\beta_1}}-\frac{(1-\gamma)^{\beta_1}(1+\beta_1\gamma)-2-\beta_1\gamma}{\gamma(1-(1-\gamma)^{\beta_1})}\\
         &= \beta_1+\frac{1+(2+\beta_1\gamma)(1-\gamma)^{\beta_1}-(1-\gamma)^{2\beta_1}(1+\beta_1\gamma)}{\gamma(1-\gamma)^{\beta_1}(1-(1-\gamma)^{\beta_1})}.
  \end{align*}
\end{proof}
This expression is complex to
analyze.  Hence, a graphical representation of this bound is given in
Figure~\ref{fig:glacier}.  It shows the expected smallest number of transactions
to be queried by an honest party (not counting no-ops) until it can confirm
the target transaction~$T$.  The larger this gets, the more the protocol
loses liveness.  It is relevant that this bound grows proportional to
$\frac{1}{(1-\gamma)^{\beta_1}}$, i.e., exponential in acceptance
threshold~$\beta_1$ since $(1-\gamma)<1$.

The Avalanche team has acknowledged our findings and the vulnerability in
the abstract protocol.  The protocol deployed in the actual network,
however, differs from our formalization in a way that should prevent the
problem.  We describe the deployed version of Avalanche in
Appendix~\ref{appendix:fix}.  The next section describes another variation
of Avalanche that provably eliminates the problem.

\section{Fixing liveness with Glacier}
\label{app:glacier}

The adversary is able to delay the acceptance of an honest transaction~$T$
because $T$ is directly influenced by the queries of its descendants.  Note
the issuer of $T$ has no control over its descendants according to the
protocol.  A unsuccessful
query of a descendant
of $T$ carries a negative consequence for the acceptance of $T$, regardless
of the status of $T$ inside its conflict set. This influence is the root of
the problem described earlier.
An immediate, but inefficient remedy might be to run one Snowball consensus
instance for each transaction. However, this would greatly degrade the
throughput and increase the latency of the protocol, as many more
messages would be exchanged.

We propose here a modification, called \emph{Glacier}, in which an
unsuccessful query of a transaction~$T$ carries negative consequences
\emph{only} for those of its ancestors that led to negative votes and
caused the query to be unsuccessful. Our protocol is shown in
Algorithm~\ref{algo:glacier}. It specifically modifies the voting protocol
and adds to each \str{Vote} message for $T$ a list~$L$ with all ancestors
of $T$ that are not preferred in their respective conflict sets
(L~\ref{l:ancestors}--\ref{l:voteancestor}).
When party~$u$ receives a negative vote like $\msg{Vote}{v, T, \false, L}$,
it performs the same actions as before.  Additionally, it increments a
counter for each ancestor $T^\ast$ of $T$ to denote how many parties have
reported $T^\ast$ as not preferred while accepting~$T$ (L~\ref{l:addnonpref}).
If $u$ receives a positive vote, the protocol remains unchanged.

If the query is successful because $u$ receives at least $\alpha$ positive
votes on~$T$, then it proceeds as before (Algorithm~\ref{algo:aux2},
L~\ref{l:positive1}).  But before $u$ declares the query to be
unsuccessful, it furthermore waits until having received a vote on $T$ from
all $k$ parties sampled in the query (L~\ref{l:g_negative1}).  When this is
the case, $u$ only resets the counter for acceptance of those ancestors
$T^\ast$ of $T$ that have been reported as non-preferred by more than
$k-\alpha$ queried parties (L~\ref{l:g_ancestors}--\ref{l:g_negative2}).
If $T^\ast$ is preferred by at least $\alpha$ parties, however, then $u$
increments its confidence level as before (L~\ref{l:g_increment}).

\begin{algo*}[h]
  \begin{numbertabbing}
    xxxx\=xxxx\=xxxx\=xxxx\=xxxx\=xxxx\=xxxx\kill
    \textbf{State}\\
    \> $\var{nonpref}:\op{HashMap}[\CT\times\CT\to \mathbb{N} ]$
    \` // number of votes on $T$ reporting that $T'$ is not preferred\label{}\\
    \\[-1ex]
    \textbf{upon} receiving message \msg{Query}{T} from party $v$ \textbf{do}\` // replaces L~\ref{l:receiveTransaction}\label{}\\
    \>$L\gets[\ ]$\` // contains the non-preferred ancestors of $T$\label{l:ancestors}\\
    \>\textbf{for} $T'\in\var{ancestors}(T)$ \textbf{do}\label{}\\
    \>\>\textbf{if} $\neg\op{preferred}(T')$ \textbf{then}\label{}\\
    \>\>\> append $T'$ to $L$\label{}\\
    \>send message \msg{Vote}{v,T,\op{stronglyPreferred($T$)},L} to party $v$\label{l:voteancestor}\\
    \\[-1ex]
    // replaces code at L~\ref{l:receiveVote}\label{}\\
    \textbf{upon} receiving message \msg{Vote}{v,T,w, L} from a party $v\in\CS[T]$ \textbf{do}\` // $w$ is the vote\label{}\\
    \>$\op{votes}[T,v]\gets w$\label{}\\
    \>  \textbf{for} $T' \in L$ \textbf{do}\label{}\\
    \>\> \textbf{if} $\var{nonpref}[T,T']=\perp$ \textbf{then}\label{}\\
    \>\>\> $\var{nonpref}[T,T']\gets 1$\label{}\\
    \>\> \textbf{else}\label{}\\
    \>\>\> $\var{nonpref}[T,T']\gets \var{nonpref}[T, T']+ 1$\label{l:addnonpref}\\
    \\[-1ex]
    // replaces code at L~\ref{l:negative1}\label{}\\
    \textbf{upon} $\exists T \in \CT$ such that $\bigl| \op{votes}[T,v] \bigr|=k \land \Bigl| \bigl\{v \in \CS[T] \mid \op{votes}[T, v] = \false \bigr\}\Bigr| > k - \alpha$ \textbf{do}\label{l:g_negative1}\\
    \> \textbf{stop} $\op{timer}[T]$\label{}\\
     \> $\op{votes}[T, *] \gets \bot$\` // remove all entries in \op{votes} for $T$\label{}\\ 
    \> $\var{S}[T]\gets[\ ]$\` // reset the HashMap $S$\label{}\\ 
    \> \textbf{for} $T'$ such that $\var{nonpref}[T,T'] \neq \bot$ \textbf{do}\` // all ancestors of $T$\label{l:g_ancestors}\\
    \>\> \textbf{if} $\var{nonpref}[T,T']>k-\alpha$ \textbf{then}\label{}\\
    \>\>\>$\op{cnt}[\var{conflictSet}[T']]\gets 0$\label{l:g_negative2}\\
    \>\> \textbf{else} // $\var{nonpref}[T,T'] \leq \alpha$\label{}\\
    \>\>\>$\op{cnt}[\var{conflictSet}[T']]\gets \op{cnt}[\var{conflictSet}[T']]+1$\label{l:g_increment}\\
    \>$\var{nonpref}[T, \ast] \gets \bot$\label{}%
  \end{numbertabbing}
  \caption{Modifications to Avalanche (Algorithm~\ref{algo:state}--\ref{algo:aux3}) for Glacier (party $u$)}
  \label{algo:glacier}
\end{algo*}

Considering the adversary introduced in Section~\ref{sec:shortliveness}, a negative
reply to the query of a descendant of the target transaction $T$ does not carry any negative consequence for the acceptance
of $T$ here. In particular, the counter for acceptance of transaction $T$ is never reset, even when a query is unsuccessful, because
$T$ is the only transaction in its conflicting set, then always preferred. Thus, transaction $T$ will be accepted after $\beta_1$ successful queries,
if all its parents are accepted, or $\beta_2$ successful queries if they are not accepted. Assuming that queries of honest transactions are successful, on average $\frac{\beta}{1-p}$ transactions are required to accept $T$ for $\beta\in[\beta_1,\beta_2]$ depending on the state of the parents of $T$. For simplicity we assume that the parents are accepted, thus, the counter needs to achieve the value $\beta_1$. If this were not the case, then it is sufficient to substitute $\beta_1$ with $\beta$ in the upcoming expression. 
Avalanche requires on average $\beta_1+\frac{1+(2+\beta_1\gamma)(1-\gamma)^{\beta_1}-(1-\gamma)^{2\beta_1}(1+\beta_1\gamma)}{\gamma(1-\gamma)^{\beta_1}(1-(1-\gamma)^{\beta_1})}$ transactions to accept $T$ by Lemma~\ref{lemma:malicious}.
The assumption that the query of honest transactions is always successful is more beneficial to Avalanche than to Glacier, since in Avalanche such a query resets the counter for acceptance of $T$. But in Glacier, the query simply leaves the counter as it is. The value of the acceptance threshold $\beta_1$ is also more beneficial for Avalanche since the number of required transactions increases linearly in Glacier and exponentially in Avalanche. Figure~\ref{fig:glacier} shows a comparison of both expressions.

  In Glacier, the vote for a transaction is independent of the vote of its
  descendant and ancestors, even if a query of a transaction carries an
  implicit query of all its ancestors. Thus, Lemma~\ref{counters} can be extended.

\begin{lemma}\label{counters2}
  Party $u$ delivers a transaction $T$ with counter for acceptance 
  $\op{cnt}[\op{conflictSet}[T]] \geq \beta_1$ in Glacier if and
  only if $u$ decides 1 in the equivalent execution of 
  Snowball with threshold $\op{cnt}[\op{conflictSet}[T]]$.
\end{lemma}

\begin{proof}
  Consider a transaction $T$ in the equivalent execution of Snowball.
  The counter for acceptance of the value 1 in Snowball
  is always the same as the counter for acceptance of transaction $T$
  in Glacier because of the modifications introduced by Glacier. 
  Thus, following the same argument as in Lemma~\ref{counters},
  transaction $T$ is accepted in Glacier with counter $\op{cnt}[\op{conflictSet}[T]]$ if
  and only if 1 is decided with counter $\op{cnt}[\op{conflictSet}[T]]$ in the equivalent
  execution of Snowball.
\end{proof}

\begin{theorem}
  The Glacier algorithm satisfies the properties of generic broadcast in
  the presence of an adversary that controls up to $\mathcal{O}(\sqrt{n})$
  parties.
\end{theorem}

\begin{proof}
  Lemma~\ref{counters} is a a special case of Lemma~\ref{counters2}.
  Theorem~\ref{theo:reduction} shows that Lemma~\ref{counters}
  and the properties of Snowball (Appendix~\ref{sec:snowball})
  guarantee that Avalanche satisfies integrity, partial order, and 
  external validity. In the same way, Lemma~\ref{counters2} guarantees
  that Glacier satisfies these same properties.
  Thus, it is sufficient to prove that Glacier satisfies validity and agreement.
  
  \begin{description}
  \item[Validity.] Assume that an honest party broadcasts a payload $\var{tx}$.
  Because the party is honest, the transaction $T$ containing $\var{tx}$ is valid
  and non-conflicting. In the equivalent execution of Snowball, every honest
  party that proposes a value proposes 1. Hence, using the validity and termination
  properties of Snowball, every honest party eventually decides 1. Using Lemma~\ref{counters2},
  every honest party eventually delivers $\var{tx}$.
  \item[Agreement.] Assume that an honest party delivers a payload transaction $\var{tx}$ contained
  in transaction $T$. Using Lemma~\ref{counters2}, an honest party decides 1 in the equivalent
  execution of Snowball. Because of the termination and agreement
  properties of Snowball, every honest party decides $1$. Using Lemma~\ref{counters2} again,
  every honest party eventually delivers payload $\var{tx}$
  \end{description}
We conclude that Glacier satisfies the properties of generic broadcast.
\end{proof}

With the modification to Glacier, Avalanche can be safely used as the basis for a payment system.
The only possible concern with Glacier could be a decrease in performance compared to Avalanche.
However, Glacier does not reduce the performance but rather improves it.
Glacier only modifies the update in the local state of party $u$ after a query has been unsuccessful.
The counter of acceptance of a given transaction $T$ in Glacier implementation is always 
greater or equal than its counterpart in Avalanche. This follows because a reset of 
$\op{cnt}[\var{conflictSet}[T]]$ in Glacier implies the same reset in Avalanche. Such a reset
in Glacier occurs if the query of a descendant of $T$ fails and $T$ was reported as non-preferred by 
more than $k-\alpha$ parties, whereas in Avalanche it is enough if the query of the descendant failed.
In Avalanche, $\op{cnt}[\var{conflictSet}[T]]$ is incremented if the query of a descendant of $T$ succeeds, and
the same occurs in Glacier. Thus, $\op{cnt}[\var{conflictSet}[T]]$ in Glacier is always greater or equal than
in Avalanche. We recall that a transaction is accepted when $\op{cnt}[\var{conflictSet}[T]]$ reaches a
threshold depending on some conditions of the local view of the DAG, but these are identical for
Glacier and Avalanche. Hence, every transaction that is accepted in Avalanche is accepted in Glacier with equal
or smaller latency. This implies not only that the latency of Glacier is smaller than the latency of Avalanche, but
also that the throughput of Glacier is at least as good as the throughput of Avalanche.

\section{Conclusion}

Avalanche is well-known for its remarkable throughput and latency that are achieved through a metastable sampling technique. The pseudocode we introduce captures in a compact and relatively simple manner the intricacies of the system. We show that Avalanche, as originally introduced, possesses a vulnerability allowing an adversary to delay transactions arbitrarily. We also address such vulnerability with a modification of the protocol, Glacier, that allows Avalanche to satisfy both safety and liveness.

The developers of Avalanche have acknowledged the vulnerability, and the actual implementation does not suffer from it due to an alternative fix. Understanding this variant of Avalanche remains open and is subject of future work.

\section*{Acknowledgments}

This work has been funded by the Swiss National Science Foundation (SNSF)
under grant agreement Nr\@.~200021\_188443 (Advanced Consensus Protocols).

\bibliographystyle{acm}
\bibliography{references,dblpbibtex}

\begin{thebibliography}{10}

\bibitem{DBLP:conf/opodis/Amores-SesarCM20}
{\sc Amores{-}Sesar, I., Cachin, C., and Micic, J.}
\newblock Security analysis of ripple consensus.
\newblock In {\em {OPODIS}\/} (2020), vol.~184 of {\em LIPIcs}, Schloss
  Dagstuhl - Leibniz-Zentrum f{\"{u}}r Informatik, pp.~10:1--10:16.

\bibitem{DBLP:conf/trust/ArmknechtKMYZ15}
{\sc Armknecht, F., Karame, G.~O., Mandal, A., Youssef, F., and Zenner, E.}
\newblock Ripple: Overview and outlook.
\newblock In {\em {TRUST}\/} (2015), vol.~9229 of {\em Lecture Notes in
  Computer Science}, Springer, pp.~163--180.

\bibitem{documentation}
{\sc {Ava Labs, Inc.}}
\newblock Avalanche documentation.
\newblock \url{https://docs.avax.network/}.

\bibitem{source}
{\sc {Ava Labs, Inc.}}
\newblock Node implementation for the {Avalanche} network.
\newblock \url{https://github.com/ava-labs/avalanchego}.

\bibitem{DBLP:books/daglib/0025983}
{\sc Cachin, C., Guerraoui, R., and Rodrigues, L. E.~T.}
\newblock {\em Introduction to Reliable and Secure Distributed Programming
  {(2.} ed.)}.
\newblock Springer, 2011.

\bibitem{DBLP:conf/crypto/CachinKPS01}
{\sc Cachin, C., Kursawe, K., Petzold, F., and Shoup, V.}
\newblock Secure and efficient asynchronous broadcast protocols.
\newblock In {\em {CRYPTO}\/} (2001), vol.~2139 of {\em Lecture Notes in
  Computer Science}, Springer, pp.~524--541.

\bibitem{DBLP:conf/wdag/CachinV17}
{\sc Cachin, C., and Vukolic, M.}
\newblock Blockchain consensus protocols in the wild (keynote talk).
\newblock In {\em {DISC}\/} (2017), vol.~91 of {\em LIPIcs}, Schloss Dagstuhl -
  Leibniz-Zentrum f{\"{u}}r Informatik, pp.~1:1--1:16.

\bibitem{coinmarketcap}
Coinmarketcap: Today's cryptocurrency prices by market cap.
\newblock \url{https://coinmarketcap.com/}, 2022.

\bibitem{DBLP:conf/eurocrypt/DavidGKR18}
{\sc David, B., Gazi, P., Kiayias, A., and Russell, A.}
\newblock Ouroboros praos: An adaptively-secure, semi-synchronous
  proof-of-stake blockchain.
\newblock In {\em {EUROCRYPT} {(2)}\/} (2018), vol.~10821 of {\em Lecture Notes
  in Computer Science}, Springer, pp.~66--98.

\bibitem{DBLP:journals/jacm/DworkLS88}
{\sc Dwork, C., Lynch, N.~A., and Stockmeyer, L.~J.}
\newblock Consensus in the presence of partial synchrony.
\newblock {\em J. {ACM} 35}, 2 (1988), 288--323.

\bibitem{DBLP:conf/fc/EyalS14}
{\sc Eyal, I., and Sirer, E.~G.}
\newblock Majority is not enough: Bitcoin mining is vulnerable.
\newblock In {\em Financial Cryptography\/} (2014), vol.~8437 of {\em Lecture
  Notes in Computer Science}, Springer, pp.~436--454.

\bibitem{DBLP:conf/eurocrypt/GarayKL15}
{\sc Garay, J.~A., Kiayias, A., and Leonardos, N.}
\newblock The bitcoin backbone protocol: Analysis and applications.
\newblock In {\em {EUROCRYPT} {(2)}\/} (2015), vol.~9057 of {\em Lecture Notes
  in Computer Science}, Springer, pp.~281--310.

\bibitem{DBLP:conf/crypto/GarayKL17}
{\sc Garay, J.~A., Kiayias, A., and Leonardos, N.}
\newblock The bitcoin backbone protocol with chains of variable difficulty.
\newblock In {\em {CRYPTO} {(1)}\/} (2017), vol.~10401 of {\em Lecture Notes in
  Computer Science}, Springer, pp.~291--323.

\bibitem{DBLP:conf/sosp/GiladHMVZ17}
{\sc Gilad, Y., Hemo, R., Micali, S., Vlachos, G., and Zeldovich, N.}
\newblock Algorand: Scaling byzantine agreements for cryptocurrencies.
\newblock In {\em {SOSP}\/} (2017), {ACM}, pp.~51--68.

\bibitem{DBLP:conf/podc/KeidarKNS21}
{\sc Keidar, I., Kokoris{-}Kogias, E., Naor, O., and Spiegelman, A.}
\newblock All you need is {DAG}.
\newblock In {\em {PODC}\/} (2021), {ACM}, pp.~165--175.

\bibitem{DBLP:conf/crypto/KiayiasRDO17}
{\sc Kiayias, A., Russell, A., David, B., and Oliynykov, R.}
\newblock Ouroboros: {A} provably secure proof-of-stake blockchain protocol.
\newblock In {\em {CRYPTO} {(1)}\/} (2017), vol.~10401 of {\em Lecture Notes in
  Computer Science}, Springer, pp.~357--388.

\bibitem{DBLP:conf/eurosp/KimKK19}
{\sc Kim, M., Kwon, Y., and Kim, Y.}
\newblock Is stellar as secure as you think?
\newblock In {\em EuroS{\&}P Workshops\/} (2019), {IEEE}, pp.~377--385.

\bibitem{DBLP:journals/corr/abs-1805-03870}
{\sc Li, C., Li, P., Xu, W., Long, F., and Yao, A.~C.}
\newblock Scaling nakamoto consensus to thousands of transactions per second.
\newblock {\em CoRR abs/1805.03870\/} (2018).

\bibitem{DBLP:conf/usenix/LiLZYWYXLY20}
{\sc Li, C., Li, P., Zhou, D., Yang, Z., Wu, M., Yang, G., Xu, W., Long, F.,
  and Yao, A.~C.}
\newblock A decentralized blockchain with high throughput and fast
  confirmation.
\newblock In {\em {USENIX} Annual Technical Conference\/} (2020), {USENIX}
  Association, pp.~515--528.

\bibitem{DBLP:conf/sosp/LokhavaLMHBGJMM19}
{\sc Lokhava, M., Losa, G., Mazi{\`{e}}res, D., Hoare, G., Barry, N., Gafni,
  E., Jove, J., Malinowsky, R., and McCaleb, J.}
\newblock Fast and secure global payments with stellar.
\newblock In {\em {SOSP}\/} (2019), {ACM}, pp.~80--96.

\bibitem{DBLP:journals/corr/abs-2111-07805}
{\sc Mamache, H., Mazu{\'{e}}, G., Rashid, O., Bu, G., and Potop{-}Butucaru,
  M.}
\newblock Resilience of {IOTA} consensus.
\newblock {\em CoRR abs/2111.07805\/} (2021).

\bibitem{bitcoin}
{\sc Nakamoto, S.}
\newblock Bitcoin: A peer-to-peer electronic cash system.
\newblock Whitepaper, \url{https://bitcoin.org/bitcoin.pdf}, 2009.

\bibitem{DBLP:journals/jacm/PeaseSL80}
{\sc Pease, M.~C., Shostak, R.~E., and Lamport, L.}
\newblock Reaching agreement in the presence of faults.
\newblock {\em J. {ACM} 27}, 2 (1980), 228--234.

\bibitem{DBLP:conf/wdag/PedoneS99}
{\sc Pedone, F., and Schiper, A.}
\newblock Generic broadcast.
\newblock In {\em {DISC}\/} (1999), vol.~1693 of {\em Lecture Notes in Computer
  Science}, Springer, pp.~94--108.

\bibitem{ross1996stochastic}
{\sc Ross, S.}
\newblock {\em Stochastic Processes}, second~ed.
\newblock Wiley, 1996.

\bibitem{DBLP:conf/aft/SompolinskyWZ21}
{\sc Sompolinsky, Y., Wyborski, S., and Zohar, A.}
\newblock {PHANTOM} {GHOSTDAG:} a scalable generalization of nakamoto
  consensus: September 2, 2021.
\newblock In {\em {AFT}\/} (2021), {ACM}, pp.~57--70.

\bibitem{polling}
{\sc Tan, W.~Y.}
\newblock On the absorption probabilities and absorption times of finite
  homogeneous birth-death processes.
\newblock {\em Biometrics 32}, 4 (1976), 745--752.

\bibitem{avalanche}
{\sc {Team Rocket}, Yin, M., Sekniqi, K., van Renesse, R., and Sirer, E.~G.}
\newblock Scalable and probabilistic leaderless {BFT} consensus through
  metastability.
\newblock e-print, arXiv:1906.08936 [cs.CR], 2019.

\bibitem{DBLP:journals/corr/abs-2008-04863}
{\sc Wang, B., Wang, Q., Chen, S., and Xiang, Y.}
\newblock Security analysis on tangle-based blockchain through simulation.
\newblock {\em CoRR abs/2008.04863\/} (2020).

\bibitem{DBLP:conf/fc/WangYPB0D020}
{\sc Wang, Q., Yu, J., Peng, Z., Bui, V.~C., Chen, S., Ding, Y., and Xiang, Y.}
\newblock Security analysis on dbft protocol of {NEO}.
\newblock In {\em Financial Cryptography\/} (2020), vol.~12059 of {\em Lecture
  Notes in Computer Science}, Springer, pp.~20--31.

\bibitem{DBLP:phd/us/Yin21}
{\sc Yin, M.}
\newblock {\em Scaling the Infrastructure of Practical Blockchain Systems}.
\newblock PhD thesis, Cornell University, {USA}, 2021.

\end{thebibliography}
 
\appendix

\section{The Snowball protocol}
\label{sec:snowball}

The Snowball protocol is the Byzantine-resistant protocol introduced together
with Avalanche in the whitepaper~\cite{avalanche}. We shortly
summarize this protocol in Algorithm~\ref{algo:snowball}.

\begin{algo*}
  \begin{numbertabbing}
    xxxx\=xxxx\=xxxx\=xxxx\=xxxx\=xxxx\=xxxx\kill
     \textbf{Global parameters and state}\\
      $\CN$ \` // set of parties\label{}\\
      $\var{newRound}\in\{\false,\true\}$ \` // boolean variable indicating when to start a round\label{}\\
      $\var{decided}\in\{\false,\true\}$ \` // boolean variable indicating when to finish the protocol\label{}\\
  	  $k\in\mathbb{N}$  \` // number of parties queried in each poll\label{}\\
  	  $\alpha\in\mathbb{N}$ \` // majority threshold for queries\label{}\\
  	  $\var{cnt}\in\mathbb{N}$ \` // counter for acceptance\label{}\\
  	  $\beta\in\mathbb{N}$ \` // threshold for acceptance\label{}\\
      $\CS:\op{HashMap}[\CT\to\CN]$ \` // set of sampled parties to be queried\label{}\\
      $d:\op{HashMap}[\{0,1\}\to\mathbb{N}]$ \` // confidence value of a transaction\label{}\\
      $\var{votes}:\op{HashMap}[\{0,1\}\to\mathbb{N}]$ \` // number of votes for a value\label{}\\
      \\
      \textbf{Algorithm}\\
     \textbf{upon} $\op{propose}(b)$ \textbf{do}\label{}\\
     \>$\var{decided}\gets\false$\label{}\\
     \>$\var{newRound}\gets\true$\label{}\\
     \\
    	\textbf{upon} $\var{newRound}\land\neg\var{decided}$ \textbf{do}\` // still not decided\label{}\\
    	\>$\var{newRound}\gets\false$\label{}\\
    	\> \textbf{if} $b\neq\perp$ \textbf{then}\label{}\\
    \>\>$\CS \gets\op{sample}(\CN\setminus\{u\}, k)$\` // sample $k$ random parties\label{}\\
	\>\> \op{send} message \msg{Query}{b} to all parties $v\in \CS$\label{}\\    
	\\
    \textbf{upon} $\var{votes}[b']\geq \alpha$ \textbf{do}\` // $b'=0$ or $b'=1$\label{l:queryend}\\
    \>$d[b']\gets d[b']+1$\label{}\\
    \>\textbf{if} $b= b'$ \textbf{then}\` // the outcome of the query is the same as our proposal\label{}\\
    \>\> $\var{cnt}\gets \var{cnt}+1$\label{}\\
    \> \textbf{else}\label{}\\
    \>\>\textbf{if} $d[b']>d[b]$ \textbf{then}\label{}\\
    \>\>\> $b\gets b'$\label{}\\
    \>\>\> $\var{cnt}\gets 0$\label{}\\
    \> $\var{newRound}\gets\true$\label{}\\
    \\
   \textbf{upon} $n=k\land \var{votes}[0]<\alpha\land \var{votes}[1]<\alpha$ \textbf{do}\` // there is no majority for any value\label{}\\
   \> $\var{cnt}\gets 0$\label{}\\
   \>$\var{newRound}\gets\true$\label{}\\
   \\
   \textbf{upon} receiving message \msg{Query}{b'} from party $v$ \textbf{do}\label{}\\
   \>\textbf{if} $b=\perp$ \textbf{then}\label{}\\
   \>\>$\var{decided}\gets\false$\label{}\\
   \>\>$b\gets b'$\label{}\\
   \>\op{send} message \msg{Vote}{b} to party $v$\` // reply with the local value of $b$\label{}\\
   \\
   \textbf{upon} receiving message \msg{Vote}{b^\ast} from a party $v\in\CS$ \textbf{do}\` // collect the vote $b^\ast$\label{}\\
   \>$\var{votes}[b^\ast]\gets \var{votes}[b^\ast]+1$\label{}\\
   \\
   \textbf{upon} $\var{cnt}=\beta$ \textbf{do}\` // there is enough confidence for $B$\label{}\\
   \>$\op{decide}(b)$\label{}\\
   \>$\var{decided}\gets\true$\label{}
           \end{numbertabbing}
    \caption{Snowball (party $u$)}
    
   \label{algo:snowball}
  \end{algo*}

Snowball possesses almost the same structure as Avalanche, but it is a protocol for consensus, not for broadcast. 
This Byzantine consensus primitive is accessed through the events
$\var{propose}(\var{b})$ and $\var{decide}(\var{b})$. Any party
is allowed to propose a value $b\in\{0,1\}$. Since Snowball is a
probabilistic algorithm, the properties of our abstraction need to 
be fulfilled only with all but negligible probability.
  \begin{definition}\label{def:vbc}
  A protocol solves \emph{Byzantine consensus} if it satisfies the following conditions, except with negligible probability:
  \begin{description}
  \item[Validity:] If all parties are honest and \emph{propose} the same value $v$, then no honest party
  decides a value different from $v$; furthermore, if some process decides $v$, then $v$ was proposed by some process.
  \item[Termination:] Every honest party eventually \emph{decides} some value.
  \item[Integrity:] No honest party \emph{decides} twice.
  \item[Agreement:] No two honest parties \emph{decide} differently.
 \end{description}
 \end{definition}

At the beginning of the protocol, party $u$ may propose
a value $b$, if $u$ does not propose $b=\perp$. Snowball is structured in rounds around the same sample mechanism
as Avalanche. Party $u$ starts a round by sampling $k$ parties at random and querying them for the value they are 
currently considering for $b$. If the value of a queried party is undefined, the queried party adopts the value that it is been queried with and replies accordingly.
If more than $\alpha$ votes for value $b'$ are collected (L~\ref{l:queryend}), the query is finalized and $u$ updates its local state by incrementing $d[b']$. If $b'$ is the same value as the value $b$ in the local view of $u$, the counter for acceptance is incremented. However, if $b'\neq b$ and $d[b']>d[b]$, the party updates its local value $b$ and resets the counter to zero. Party $u$ finishes consensus when the acceptance counter reaches
the value $\beta$.

Considering the adversary introduced in Section~\ref{sec:adversary} controlling up to $O(\sqrt{n})$ parties, 
it can be shown that Snowball satisfies the properties of Byzantine consensus.
\begin{theorem}
Snowball satisfies Byzantine consensus.
\end{theorem}
\begin{proof}
The proof is provided in the Appendix~A of the whitepaper of Avalanche~\cite{avalanche}.
\end{proof}

\clearpage

\section{The Avalanche protocol as implemented}
\label{appendix:fix}

The actual implementation of Avalanches addresses the liveness problems differently from Glacier and works as follows. Consider the protocol in Section~\ref{sec:description}.  In the implementation, a party $u$ queries $k$ parties with transaction~$T$ as before.  When some party~$v$ is queried with $T$, then $v$ first adds $T$ to its local view.  Then it replies with a \str{Vote} message, but instead of including a binary vote, it sends the whole virtuous frontier according to its local view.  Party $u$ collects the responses as in Section~\ref{sec:description} and stores the virtuous frontiers received in the \str{Vote} messages.  Then it counts how many queried parties have reported some transaction $T'$, or a descendant of $T'$, as part of their virtuous frontier in the variable $\var{ack}[T,T']$. If more than $\alpha$ parties have reported $T'$, then $u$ adds $T'$ to the set $\CG[T']$ and updates its state, as for a successful query in the original protocol (L~\ref{l:newpositive1}--\ref{l:newpositive2}). For the remaining transactions, i.e., all the transactions in \CQ outside $\CG[T']$, the counter is set to zero (L~\ref{l:newnegative1}--\ref{l:newnegative2}); this is equivalent to the effect of a negative query in the original protocol.

This fix addresses the liveness issue shown in Section~\ref{sec:behavior} since a potential adversary loses the ability to submit a transaction that causes a reset of the acceptance counter of an honest transaction. As explained in Section~\ref{sec:adversary}, the adversary could reset the counter for acceptance of honest transactions by simply submitting a transaction $T$ referring to the target transaction $T$ and both transactions of a double spending $T_1$ and $T_2$ because $T$ is not strongly preferred. However, in the implemented protocol, the parties reply with the virtuous frontier regardless of the queried transaction. Due to this, the adversary cannot influence the reply of the queried parties by creating malicious transactions.
A detailed analysis of this protocol is beyond the scope of this work, however.

\begin{algo*}
  \begin{numbertabbing}
    xxxx\=xxxx\=xxxx\=xxxx\=xxxx\=xxxx\=xxxx\kill
    \textbf{State}\\
    $\op{ack}:\op{HashMap}[\CT\times\CT\to \mathbb{N} ]$
    \` // number of votes on $T$ reporting that $T'$ is not preferred\label{}\\
   $\CG:\op{HashMap}[\CT\times\CN\to 2^{\CT} ]$
    \` // ancestors of positively reported transactions\label{}\\
    \\[-1ex]
    // replaces code at L~\ref{l:receiveTransaction}\label{}\\
    \textbf{upon} receiving message \msg{Query}{T} from party $v$ \textbf{do}\label{}\\
    \> \textbf{if} $T\not\in\CT$ \textbf{then} \` // party $u$ sees $T$ for the first time\label{}\\
    \>\> $\op{updateDAG}(T)$ \label{}\\
    \>send message \msg{Vote}{u,T,\CVF} to party $v$\label{l:votefrontier}\\
    \\[-1ex]
    // replaces code at L~\ref{l:receiveVote}\label{}\\
    \textbf{upon} receiving message \msg{Vote}{v,T,\CVF'} from a party $v\in\CS[T]$ \textbf{do}\` // $\CVF'$ is the new vote\label{}\\
    \>$\op{votes}[T,v]\gets \CVF'$\label{}\\
    \>\textbf{for} $T'\in\op{votes}[T,v]$ \textbf{do}\` // build the ancestors of the reported transactions\label{}\\
    \>\>$\CG[T,v]\gets \CG[T,v]\cup\op{ancestors}(T')$\label{}\\
    \>\textbf{for} $T'\in\CG[T,v]$ \textbf{do}\` // count the number of parties that reported $T'$\label{}\\
    \>\>\textbf{if} $\op{ack}[T,T']=\perp$ \textbf{then}\label{}\\
    \>\>\> $\op{ack}[T,T']\gets 1$\label{}\\
    \>\> \textbf{else}\label{}\\
    \>\> $\var{ack}[T,T']\gets \var{ack}[T, T']+ 1$\label{}\\
    \\[-1ex]
    
    // replaces code at L~\ref{l:positive1} and L~\ref{l:negative1}\label{}\\
\textbf{upon} $\exists T \in \CT$ such that $\bigl|\left\{ \op{votes}[T,v] \right\}\bigr|=k$\` // every queried party has replied\label{}\\
    \> \textbf{stop} $\op{timer}[T]$\label{}\\
    \> $\op{votes}[T, *] \gets \bot$\` // remove all entries in \op{votes} for $T$\label{}\\ 

    \>\textbf{for} $T'\in\CQ$ \textbf{do}\label{}\\
    \>\>\textbf{if} $\var{ack}[T,T']\geq\alpha$ \textbf{then}\label{l:newpositive1}\\
    \>\>\> $d[T']\gets d[T']+1$\label{}\\
    \>\>\>\textbf{if} $d[T']>d[\op{pref}[\var{conflictSet}[T']]]$ \textbf{then}\label{}\\
    \>\>\>\> $\op{pref}[\var{conflictSet}[T']]\gets T'$\label{}\\
    \>\>\>\textbf{if} $T'\neq \op{last}[\var{conflictSet}[T']]$ \textbf{then}\label{l:newchange}\\
    \>\>\>\> $\op{last}[\var{conflictSet}[T']]\gets T'$\label{}\\
    \>\>\>\>$\op{cnt}[\var{conflictSet}[T']]\gets1$\label{l:newpositive2}\\
    \>\>\>\textbf{else}\label{}\\
    \>\>\>\> $\op{cnt}[\var{conflictSet}[T']]\gets \op{cnt}[\var{conflictSet}[T']]+1$\label{l:newnegative1}\\
    \>\>\textbf{else}\label{}\\
    \>\>\>$\op{cnt}[\var{conflictSet}[T']]\gets 0$\label{l:newnegative2}\\
    
	\> $\op{ack}[T, *] \gets \bot$\` // remove all entries in \op{ack} for $T$\label{}\\ 
	\> $\CG[T, *] \gets \bot$\` // remove all entries in \CG for $T$\label{}
  \end{numbertabbing}
  \caption{Modifications to Avalanche (Algorithm~\ref{algo:state}--\ref{algo:aux3}) in the implementation (party $u$)}
  \label{algo:current}
\end{algo*}

\end{document}